\documentclass{new_tlp}
\usepackage{times}
\usepackage{graphicx}
\usepackage{latexsym}
\usepackage{helvet}
\usepackage{courier}
\usepackage{amsmath}
\usepackage{amssymb}
\usepackage{algorithm}
\usepackage{alltt}
\usepackage{verbatim}
\usepackage{url}
\usepackage{xspace}
\usepackage[defblank]{paralist}

\usepackage{pgfplots}
\pgfplotsset{compat=1.14}
\usepackage{tikz}
\usepackage[caption=false]{subfig}

\usepackage{algorithmicx}
\usepackage{algpseudocode}

\newcommand{\ignore}[1]{}
\newcommand{\finish}[1]{}
\newcommand{\skipit}[1]{{ #1}}

\newcommand{\uplt}{\rotatebox[origin=c]{90}{$\mathbf{<}$}}

\newcommand{\DL}{{ DL}}
\newcommand{\WFDL}{{ WFDL}}
\newcommand{\pl}{\partial_{||}}
\newcommand{\md}[1]{-\partial #1}
\newcommand{\PD}[1]{+\Delta #1}

\newcommand{\ARROW}{\hookrightarrow}
\newcommand{\mt}[1]{\mathtt{#1}}
\newcommand{\non}{{\thicksim}}
\newcommand{\true}{{\bf true}}
\newcommand{\false}{{\bf false}}

\newcommand{\unknown}{{\bf unknown}}
\renewcommand{\imath}{i}
\newcommand{\lfp}{\mathit{lfp}}

\newcommand{\UU}{\U\U}
\newcommand{\WWF}{WF^{+s}}

\newcommand{\Datalog}{Datalog$^\neg$}

\newcounter{clause}
\def\theclause{$c$\arabic{clause}}

\newenvironment{clause}{\begin{tabbing}
xxx\=xxx\=xxx\=\+\kill}%
{\end{tabbing}}

\newenvironment{Clause}{\refstepcounter{clause}%
\begin{tabbing}
cxxxx\=xxx\=xxx\=\kill
\theclause\>\+}%
{\end{tabbing}}

\newtheorem{theorem}{Theorem}
\newtheorem{lemma}[theorem]{Lemma}

\newtheorem{propn}[theorem]{Proposition}

\newtheorem{corollary}[theorem]{Corollary}
\newtheorem{example}[theorem]{Example}

\newcommand{\F}{{\cal F}}

\newcommand{\M}{{\cal M}}
\renewcommand{\P}{{\cal P}}
\newcommand{\Q}{{\cal Q}}

\newcommand{\U}{{\cal U}}

\newcommand{\V}{{\cal V}}
\newcommand{\W}{{\cal W}}

\newcommand{\NN}{\mathbb{N}}

\newcommand{\mi}[1]{\mathit{#1}}

\newcommand{\seq}[2][n]{\ensuremath{#2_{1},\dots,#2_{#1}}\xspace}

\def    \md      {\sqsupseteq}

\def    \gen     { \geq_{ -1 } \, }

\def    \gep     { \geq_{ +1 } \, }

\def    \mdn     { \md_{ -1 } \, }
\def    \mdp     { \md_{ +1 } \, }

\begin{document}

\title{Defeasible Reasoning via Datalog$^\neg$}

\author[M.J. Maher]{Michael J. Maher \\
%\institute{
Reasoning Research Institute \\
Canberra, Australia  \\
E-mail: michael.maher@reasoning.org.au
%}
}

\maketitle
\bibliographystyle{acmtrans}

\begin{abstract}
We address the problem of compiling defeasible theories to \Datalog{} programs.
We prove the correctness of this compilation, for the defeasible logic $\DL(\pl)$,
but the techniques we use apply to many other defeasible logics.
Structural properties of $\DL(\pl)$ are identified that support efficient implementation and/or approximation
of the conclusions of defeasible theories in the logic, compared with other defeasible logics.
We also use previously well-studied structural properties of logic programs
to adapt to incomplete \Datalog{} implementations.

Under consideration in Theory and Practice of Logic Programming (TPLP).
\finish{remove finish and ignore, and appendices}

\end{abstract}

% De Bortoli: distributed ASP
% webpie: Map-reduce applied to RDFS and OWL


\section{Introduction}

A problem faced by defeasible logics -- among other logical languages --
is that changing hardware and software architectures are not reflected in implementations.
Hardware architectures can range
from the use of GPUs and other hardware accelerators,
through multi-core multi-threaded architectures,
to shared-nothing cloud computing.
Causes for failure to exploit these architectures include 
lack of expertise in the architectural features, 
lack of manpower more generally,
and difficulty in updating legacy systems. 
Such problems can be ameliorated by mapping a logic to logic programming as an intermediate language.

This is a common strategy in the implementation of defeasible logics.
The first implementation of a defeasible logic, d-Prolog, was implemented as a Prolog meta-interpreter \cite{Nute_subsumption}.
Courteous Logic Programs \cite{Grosof97} and its successors LPDA \cite{LPDA}, Rulelog \cite{Rulelog}, Flora2 \cite{Flora2},
are implemented in XSB \cite{XSB}.
\footnote{Not all implementations of defeasible logics use this approach.
Delores \cite{MRABM} and SPINdle \cite{Spindle} are implemented in imperative languages,
Phobos \cite{phobos} and Deimos \cite{ECAI00} are implemented in Haskell,
while DR-DEVICE \cite{DR-DEVICE} and Situated Courteous Logic Programs \cite{SweetJess} are built on rule systems.
}
The advantages of this approach are that:
1) the target language is a high-level language with many features of the logic to be implemented;
2) the problem of optimizing the implementation to take advantage of the architecture of the underlying hardware is delegated to the implementation of logic programming;
3) flexibility and portability are consequently enhanced, 
since there are multiple implementations of logic programming,
based on differing architectures;
and
4) the design and optimization of a specific logic can proceed at a higher level of abstraction.

In this paper we address a defeasible logic designed for large scale reasoning \cite{sdl}
and the compilation of function-free defeasible theories into logic programming -- 
more specifically, \Datalog{} under the well-founded semantics \cite{WF91}.
There is a multitude of implementations of variations of Datalog, coming from different motivations,
not all suitable for implementing a defeasible logic, and not all supporting traditional syntax.
In addition to implementations developed in the database \cite{IRIS,RedFox,MyriaL,LogicBlox,push,BigDatalog,RecStep,Dedalus,Datomic,Cascalog}
and logic programming \cite{XSB,YAP,IDP,clingo,DLV,lin_alg,embedded,GPU_datalog,WFS_X10,WFS_bigdata}
communities,
implementations have been developed to service the
programming language analysis \cite{Souffle,Flix,QL,muZ,bddbddb,Formulog,Ldat},
graph processing \cite{Socialite,Graspan,EmptyHeaded},
and artificial intelligence \cite{Dyna,Filardo_thesis,Vadalog,VLog,Yedalog} communities.
These implementations address a wide range of architectures,
and many provide extension beyond traditional Datalog.

However, the number of currently available implementations that provide complete support
for the well-founded semantics of  \Datalog{} is quite few.
This leads us to use structural properties of the compiled program to establish when
an incomplete implementation of  \Datalog{} can be used to provide a complete implementation
of a defeasible theory for the defeasible logic.
Specifically, we establish how properties of the initial defeasible theory
and properties of the defeasible logic
are reflected in the compiled program and combine to support complete execution of the compiled program
using incomplete implementations of  \Datalog{} (with respect to the well-founded semantics).
We also identify methods to obtain sound approximations to the conclusions of the defeasible theory
using sound but incomplete \Datalog{} implementations.

The  compilation builds on existing work.
We represent the scalable defeasible logic as a metaprogram, in the style of \cite{flexf}.
We then use a series of unfold and fold transformations 
\cite{TS} to convert the metaprogram, applied to the defeasible theory,
to a specialized logic program.
The result is encapsulated as a mapping from defeasible theories to \Datalog{} programs,
which is established as correct as a consequence of the correctness of the individual transformations.
The size of the resulting program is linear in the size of the defeasible theory.

The remainder of the paper is structured as follows.
Sections \ref{sec:LP} and \ref{sec:defeasible_logic} introduce
the necessary concepts from logic programming and defeasible reasoning.
Section \ref{sec:sdl} defines the defeasible logic $\DL(\pl)$ from \cite{sdl} while
Section \ref{sec:metaprogram} defines the corresponding metaprogram
and establishes some of its properties.
Section \ref{sec:trans} compiles the metaprogram to a simpler form using fold/unfold transformations,
while Section \ref{sec:semantics} establishes the correctness of the metaprogram with respect to
the original, proof-theoretic definition of $\DL(\pl)$.
Structural properties of the compiled program are established in Section \ref{sec:properties}.
Section \ref{sec:opt} then identifies how these properties can be used to establish
the correctness of implementations and approximations using incomplete \Datalog{} systems.

\section{Logic Programming}  \label{sec:LP}

We introduce the elements of logic programming that we will need.
The first part defines notation and terminology for syntactic aspects of logic programs,
including dependency relations.
The second part defines the semantics of logic programs we will use.

\subsection{Syntax and Structure of Logic Programs}

Let $\Pi$ be a set of predicate symbols, $\Sigma$ be a set of function symbols,
and $\V$ be a set of variables.
Each symbol has an associated arity greater or equal to 0.
A function symbol of arity 0 is called a \emph{constant},
while a predicate of arity 0 is called a \emph{proposition}.
The \emph{terms} are constructed inductively in the usual way:
any variable or constant is a term;
if $f \in \Sigma$ has arity $n$ and $t_1, \ldots, t_n$ are terms then $f(t_1, \ldots, t_n)$ is a term;
all terms can be constructed in this way.
An \emph{atom} is constructed by applying a predicate $p \in \Pi$ of arity $n$ to $n$ terms.
A \emph{literal} is either an atom or a negated atom $not~ A$, where $A$ is an atom.

A {\em logic program} 
is a collection of {\em clauses} of the form
$$A ~\mbox{:-}~ B_1 , \ldots , B_m ,
not~ C_1 , \ldots , not~ C_n$$
where 
$A , B_1 , \ldots , B_m, C_1 , \ldots , C_n$
are atoms ($m \geq 0 , n \geq 0$).
The positive literals and the negative literals
are grouped separately
purely for notational convenience.
$A$ is called the \emph{head} of the clause and the remaining literals form the \emph{body}.
The set of all clauses with predicate symbol $p$ in the head
are said to be the clauses defining $p$.
We use \emph{ground} as a synonym for variable-free.
The set of all variable-free instances of clauses in a logic program $P$
is denoted by $ground(P)$.
$ground(P)$ can be considered a propositional logic program, but it is generally infinite.
For the semantics we are interested in, $P$ and $ground(P)$ are equivalent
with respect to inference of ground literals.

A logic program $P$ is \emph{range-restricted} if every variable in the head of a clause
also appears in a positive body literal.
$P$ is \emph{negation-safe} if, for every clause, 
every variable in a negative body literal also appears in a positive body literal.
$P$ is \emph{safe} (or \emph{allowed}) if every variable in a clause
also appears in a positive body literal of that clause.
Equivalently, $P$ is safe if it is range-restricted and negation-safe.
Safety is a property that ensures domain independence \cite{AHV},
but also it can simplify the execution of a logic program.
Range-restriction guarantees that all inferred atoms are ground;
data structures and algorithms do not need to address general atoms.
Safety ensures, in addition, that if negative literals are evaluated only after the positive part of the clause
then, again, only ground atoms must be treated.
These points apply to both top-down and bottom-up execution.

A \emph{\Datalog{} program} is a logic program
where the only terms are constants and variables.

We present some notions of dependence among predicates that are derived purely from
the syntactic structure of a logic program $P$.
We follow the definitions and notation of \cite{Kunen89}.
$p$, $q$ and $r$ range over predicates.
We define
$p \mdp q$
if $p$ appears in the head of a rule and $q$
is the predicate of a positive literal in the body of that rule.
$p \mdn q$
if $p$ appears in the head of a rule and $q$
is  the predicate of a negative literal in the body of that rule.

$p \md q$ iff $p \mdp q$ or $p \mdn q$.
We say $p$ \emph{directly depends} on $q$.
$\ge$ is the transitive closure of $\md$.
If $p \ge q$ we say $p$ \emph{depends} on $q$.
$p \approx q$ iff $p \geq q$ and $q \geq p$, expressing that $p$ and $q$ are mutually recursive.
$p > q$ iff $p \geq q$ and not $q \geq p$.
If, for all $p$ in $P$, $q \geq p$, then we say $q$ is a \emph{$\geq$-largest} predicate.
$\gep$ and $\gen$ are defined inductively as the least relations such that
$p \gep p$,
and
$p \md_i q \mbox{ \em and }
q \geq_j r \mbox{ \em implies }
p \geq_{ i \cdot j } r$,
where $i \cdot j$ denotes multiplication of $i$ and $j$.
Essentially $\gep$ denotes a relation of
dependence through an even number of negations
and $\gen$ denotes
dependence through an odd number of negations.
As is usual, we will write $p \leq q$ when $q \geq p$,
and similarly for the other relations.
$\geq_0$ denotes the transitive closure of $\mdp$.

A program $P$ is \emph{stratified} if for no predicates $p$ and $q$ in $\Pi$ does
$p \approx q$ and $p \gen q$.
Let a \emph{predicate-consistent mapping} be a function that maps atoms to the non-negative integers such that, 
for every predicate, all atoms involving that predicate are mapped to the same value.
Then, alternatively, $P$ is stratified if 
there is a predicate-consistent mapping $m$ that,
for every clause like the one above,
$m(A) \geq m(B_i)$, for $1 \leq i \leq n$
and
$m(A) > m(C_j)$, for $1 \leq j \leq m$.
The $i^{th}$ \emph{stratum} $P_i$ is the set of clauses in $P$ with head predicate $p$ such that $m(p) = i$.
$P$ is {\em call-consistent}
% (or {\em negative-cycle-free})
if, for no predicate $p$, does
$p \gen p$;
that is
no predicate depends negatively on itself.
Clearly, any stratified program is call-consistent.
$P$ is \emph{hierarchical} if for no predicates $p$ and $q$, does
$p \md q$ and $q \ge p$,
that is, no predicate symbol depends on itself.
Equivalently, $P$ is hierarchical if there is a predicate-consistent mapping $m$ that,
for every clause like the one above,
$m(A) > m(B_i)$ and $m(A) > m(C_j)$.
Every hierarchical program is stratified.

A set $\P \subseteq \Pi$ of predicates in a program $P$ is \emph{downward-closed} if
whenever $p \in \P$ and $q \leq p$ then $q \in \P$.
$\P$ is  \emph{downward-closed with floor} $\F$
if $\F \subset \P$ and both $\P$ and $\F$ are downward-closed.
A \emph{signing} for $\P$ and $P$ is a function $s$ that maps $\Pi$ to $\{-1, +1\}$
such that, for $p, q \in \P$, $p \leq_i q$ implies $s(p) = s(q) \cdot i$.
A signing is extended to atoms by defining $s( p(\vec{a}) ) = s(p)$.
For any signing $s$ for a set of predicates $\P$, there is an inverted signing $\bar{s}$ defined by
$\bar{s}(p) = - s(p)$.
$s$ and $\bar{s}$ are equivalent in the sense that they divide $\P$ into the same two sets.
Obviously, $\bar{\bar{s}} = s$.
A program $P$ is said to be \emph{strict} if no predicate $p$ depends both positively and negatively
on a predicate $q$, that is, we never have $p \gep q$ and $p \gen q$.
Let all the predicates of $P$ be contained in $\P$.
If $\P$ has a signing then $P$ is strict;
if $P$ has a $\geq$-largest predicate and $P$ is strict, then $\P$ has a signing \cite{Kunen89}.

A similar set of dependencies over ground atoms can be defined by
applying these definitions to $ground(P)$.

Given a program $P$,
an infinite sequence of atoms $\{ q_i(\vec{a_i}) \}$ is \emph{unfounded} wrt a set of predicates $\Q$ if, 
for every $i$, $q_i \mdp q_{i+1}$ and $q_i \in \Q$.
We say a predicate $p$ \emph{avoids negative unfoundedness} 
wrt a signing $s$ on $\Q$ if
for every negatively signed predicate $q$ on which $p$ depends,
no $q$-atom starts an unfounded sequence wrt $\Q$.

\subsection{Semantics of Logic Programs}

We define the semantics of interest in this paper
and identify important relationships between them.

A {\em 3-valued Herbrand interpretation}
is a mapping from ground atoms to one of three truth values:
$\true$, $\false$, and $\unknown$.
This mapping can be extended to all formulas using Kleene's 3-valued logic.

Kleene's truth tables can be summarized as follows. If $\phi$ is a
boolean combination of the atoms $\true$, $\false$, and
$\unknown$, its truth value is $\true$ iff all the possible
ways of putting in $\true$ or $\false$ for the various
occurrences of $\unknown$ lead to a value $\true$ being computed
in ordinary 2-valued logic: $\phi$ gets the value $\false$ iff
$\neg\phi$ gets the value $\true$, and $\phi$ gets the value
$\unknown$ otherwise. These truth values can be extended in the
obvious way to predicate logic, thinking of the quantifiers as
infinite disjunction or conjunction.

Equivalently, a 3-valued Herbrand interpretation $I$ can be represented as the set of literals
$\{ a ~|~ I(a) = \true \} \cup \{ not~a ~|~ I(a) = \false \}$.
This representation is used in the following definitions.
The interpretations are ordered by the subset ordering on this representation.

Some semantics are defined in terms of fixedpoints of monotonic functions over a partial order.
A function $F$ is \emph{monotonic} if $x \leq y$ implies $F(x) \leq F(y)$.
A \emph{fixedpoint} of  $F$ is a value $a$ such that $F(a) = a$.
When $F$ is monotonic on a complete semi-lattice there is a least (under the $\leq$ ordering) fixedpoint.
We use $\lfp(F)$ to denote the least fixedpoint of $F$.
When $a$ is an element of the partial order, 
$\lfp(F, a)$ denotes the least fixedpoint greater than (or equal to) $a$.

Fitting \cite{Fitting} defined a semantics for a logic program $P$ in terms of a function $\Phi_P$
mapping 3-valued interpretations, which we define as follows.
\[
\begin{array}{rcl}
\Phi_P(I) &= & \Phi^+_\P(I) \cup \neg~ \Phi^-_\P(I) \\

\Phi^+_P(I) &= &  \{a ~|~ \mbox{there is a rule } a \mbox{~:-~} B \mbox{ in } ground(P)
\mbox{ where } I (B) = \true \} \\

\Phi^-_P(I) &= & \{ a ~|~ \mbox{for every rule } a \mbox{~:-~} B \mbox{ in } ground(P)
\mbox{ with head } a, I (B) = \false \}
\end{array}
\]
where $\neg S$ denotes the set $\{ not~s ~|~ s \in S \}$.

\emph{Fitting's semantics} associates with $P$ the least fixedpoint of $\Phi_P$,
$\lfp(\Phi_P)$.
This is the least 3-valued Herbrand model of the Clark completion $P^*$ of $P$.
Thus, the conclusions justified under this semantics are those formulas that evaluate to $\true$ under all
3-valued Herbrand models of $P^*$.
Kunen \cite{Kunen87} defined a semantics that justifies as conclusions  those formulas that evaluate to $\true$ under all 3-valued models (Herbrand or not) of  $P^*$.
He showed that these are exactly the formulas that are consequences of $\Phi_P \uparrow n$ for some finite $n$.\footnote{
$\uparrow$ is defined inductively: $\Phi_P \uparrow 0 = I_\unknown$, where $I_\unknown$ is the interpretation that assigns each atom the value $\unknown$, and $\Phi_P \uparrow (k+1) =  \Phi_P (\Phi_P \uparrow k)$.
For limit ordinals $\alpha$, $\Phi_P \uparrow \alpha = \bigcup_{\beta < \alpha} \Phi_P \uparrow \beta$.
The (possibly transfinite) sequence $\Phi_P \uparrow 0, \Phi_P \uparrow 1, \ldots$ is called the Kleene sequence for $\Phi_P$.
}
When $P$ is a finite \Datalog{} program the two semantics coincide.
However, in general, when function symbols are permitted,
Kunen's semantics is computable, while other semantics, like Fittting's semantics,
the stratified semantics,
and the well-founded semantics, are not.

The \emph{stratified semantics} (or \emph{iterated fixedpoint} semantics) \cite{ABW} 
applies only when $P$ is stratified.
It is defined in stages, by building up partial models, based on the strata, until a full model is constructed.
Each stratum contains essentially a definite clause program, given that all negations refer to lower strata,
which have already been defined in the current partial model.
Let $m$ be a mapping describing a stratification.
Initially the partial model leaves all predicates undefined,
and at each stratum, in turn, it is extended to define all the predicates defined in that stratum.
On each stratum $i$, the predicates on lower strata have been defined and,
because $P$ is stratified,
there is a least partial model extending the current partial model that defines the predicates.
This model is then used as the basis for the next stratum.
For a more precise description, see \cite{ABW,AptBol}.
The stratified semantics extends Fitting's semantics, when the program is stratified.

The well-founded semantics \cite{WF91} extends Fitting's semantics by, roughly,
considering atoms to be false if they are supported only by a ``loop'' of atoms.
This is based on the notion of unfounded sets.

Given a logic program $P$
and a 3-valued interpretation $I$, 
a set $A$ of ground atoms is an \emph{unfounded set
with respect to} $I$ iff each atom $a\in A$ satisfies the
following condition: For each rule $r$ of $ground(P)$
whose head is $a$, (at least) one of the following holds:
\begin{enumerate}
 \item Some literal in the body evaluates to $\false$ in $I$.
 \item Some atom in the body occurs in $A$ 
\end{enumerate}

The \emph{greatest unfounded set} of $P$ with respect to
$I$ (denoted $U_{P}(I)$) is the union of all the unfounded sets with respect to $I$.
Notice that, if we ignore the second part of the definition of unfounded set wrt $I$,
the definition of unfounded set is the same as the expression inside the definition of $\Phi^-_P(I)$.
It follows that $\Phi^-_P(I) \subseteq \U_{P}(I)$, for every $I$.

The function $\W_{P}(I)$ is  defined by
$\W_{P}=\Phi^+_P(I)  \cup \neg ~ \U_{P}(I)$.
The \emph{well-founded semantics} of a program $P$ is represented by the least fixedpoint of $W_{P}$.
This is a 3-valued Herbrand model of $P^*$.
Because $\Phi^-_P(I) \subseteq \U_{P}(I)$, for every $I$, we have $\Phi_P(I) \subseteq \W_{P}(I)$, for every $I$
and hence $\lfp(\Phi_P) \subseteq \lfp(\W_P)$.
That is, Fitting's semantics is weaker than the well-founded semantics.
If $P$ is stratified then the well-founded semantics is a 2-valued Herbrand model of $P^*$ 
and equal to the stratified semantics.

Often, as in this paper, we are mainly interested in the positive literals of predicate(s) $p$
that are consequences of the program, rather than the negative literals of $p$
or the literals of other predicates.
In such cases, we can avoid computing parts of the well-founded model \cite{signings}.
In particular,
given a signing $s$ in which $s(p)=+1$,
the positive literals of $p$ depend on only the positive literals of predicates with a positive sign,
and the negative literals of predicates with a negative sign.
Let $\WWF$ denote the semantics 
$\{ q ~|~ s(q) = +1, q \in \lfp(\W_P) \} \cup \{ not~q ~|~ s(q) = -1, not~q \in \lfp(\W_P) \}$
which computes only such parts of the well-founded model.
This can be computed as the least fixedpoint of a function $\UU_P^s$ \cite{signings}.

For a ground literal $q$,
we define: 
\begin{itemize}
\item
$P \models_{WF} q$ iff $q \in \lfp(\W_P)$
\item
$P \models_{F} q$ iff $q \in \lfp(\Phi_P)$
\item
$P \models_{\WWF} q$ iff $q \in \lfp(\UU_P^s)$
\end{itemize}
It follows from the discussion above that $P \models_{F} q$ implies $P \models_{WF} q$,
and
$P \models_{\WWF} q$ implies $P \models_{WF} q$,
for every program $P$ and ground literal $q$.

Let $I$ be a 3-valued Herbrand interpretation of predicates in a set $\F$ and 
let $X$ be a semantics based on Herbrand models.
Then $\models_X^I$ denotes 
consequence in the semantics $X$ after all predicates in $\F$ are interpreted according to $I$.
Such a notion is interesting, in general, because some predicates might be defined outside
the logic programming setting, or in a different module.

We now summarise two results from \cite{signings} in the following theorem.
In the first part we see that
only the positive or only the negative conclusions for each predicate need to be computed.
The second part establishes conditions under which the well-founded semantics and Fitting semantics agree
on the truth value of some ground literals
(even if they may disagree on other literals).

\begin{theorem}[\cite{signings}]    \label{thm:comb}
Let $P$ be a logic program, 
$\P \subseteq \Pi$ be a downward-closed set of predicates with floor $\F$,
let $\Q$ be  $\P\backslash\F$, and
$s$ be a signing for $\Q$.
Let $I$ be a fixed semantics for $\F$ and  $p \in \Q$.

\begin{enumerate}
\item   \label{thm:new}
For any ground atom $p(\vec{a})$:

If $s(p) = +1$ then
$P \models_{WF}^{I} p(\vec{a})$ ~~~~~~~~ iff ~~ $P \models_{\WWF}^{I} p(\vec{a})$

If $s(p) = -1$ then
$P \models_{WF}^{I} not~ p(\vec{a})$ ~ iff ~~ $P \models_{\WWF}^{I} not~ p(\vec{a})$

\item   \label{thm:EQ}
Suppose, additionally, that  $p$ avoids negative unfoundedness wrt $s$.

\noindent
For any ground atom $p(\vec{a})$:

If $s(p) = +1$ then
$P \models_{WF}^{I} p(\vec{a})$ ~~~~~~~~ iff ~~ $P \models_{F}^{I} p(\vec{a})$

If $s(p) = -1$ then
$P \models_{WF}^{I} not~ p(\vec{a})$ ~ iff ~~ $P \models_{F}^{I} not~ p(\vec{a})$
\end{enumerate}
\end{theorem}

\section{Defeasible Logics}  \label{sec:defeasible_logic}

A \emph{defeasible theory} $D$ is a triple $(F,R,>)$ where $F$ is a finite set of facts (literals), 
$R$ a finite set of labelled rules,
and $>$ a superiority relation (a binary acyclic relation) on $R$ (expressed on the labels),
specifying when one rule overrides another, given that both are applicable.

A rule $r$ consists (a) of its antecedent (or body) $A(r)$ which is a finite set of literals, (b) an arrow, and, (c) its
consequent (or head) $C(r)$ which is a literal. Rules also have distinct \emph{labels} which are used to refer to
the rule in the superiority relation. There are three types of rules: strict rules, defeasible rules and
defeaters represented by a respective arrow $\rightarrow$, $\Rightarrow$ and $\leadsto$. Strict rules are rules in
the classical sense: whenever the premises are indisputable (e.g., facts) then so is the conclusion. Defeasible rules
are rules that can be defeated by contrary evidence. Defeaters are rules that cannot be used to draw any conclusions; their only use is to provide contrary evidence that may prevent some conclusions.
We use $\ARROW$ to range over the different kinds of arrows used in a defeasible theory.
Given a set $R$ of rules, we denote the set of all strict rules in $R$ by
$R_{s}$, and the set of strict and defeasible rules in $R$ by $R_{sd}$. $R[q]$ denotes the set of rules in $R$ with consequent $q$.
If $q$ is a literal, $\non q$ denotes the complementary literal (if $q$ is a positive literal $p$ then $\non q$ is $\neg p$;
and if $q$ is $\neg p$, then $\non q$ is $p$).

A literal is a possibly negated predicate symbol applied to a sequence of variables and constants.
We will focus on defeasible theories such that any variable in the head of a rule also occurs in the body,
and that every fact is variable-free,
a property we call \emph{range-restricted} in analogy to the same property in logic programs.
When no rule or fact contains a function symbol, except for constants,
we say the defeasible theory is \emph{function-free}.
Given a fixed finite set of constants in a function-free defeasible theory,
any rule is equivalent to a finite set of variable-free rules, and
any defeasible theory $D$ is equivalent to a variable-free defeasible theory $ground(D)$, for the purpose of semantical analysis.
We refer to variable-free defeasible theories, etc as \emph{propositional}, since there is only a syntactic difference between such theories and true propositional defeasible theories.
Consequently, we will formulate definitions and semantical analysis in propositional terms.
However, for computational analyses and implementation we will also address defeasible theories that are not propositional.

A defeasible theory is \emph{hierarchical} (or  \emph{acyclic} or \emph{stratified}) if
there is a predicate-consistent mapping $m$ which maps atoms to the non-negative integers such that, 
for every rule,
the head is mapped to a greater value than any body atom.
That is, there is no recursion in the rules of the defeasible theory,
not even through a literal's complement.
A defeasible theory $D$ is \emph{locally hierarchical} if $ground(D)$ is hierarchical,
where we treat each variable-free atom as a proposition/0-ary predicate symbol.

\begin{example}    \label{ex:tweety}
To demonstrate defeasible theories,
we consider the familiar Tweety problem and its representation as a defeasible theory.
The defeasible theory $D$ consists of the rules and facts
\[
\begin{array}{rrcl}
r_{1}: & \mi{bird}(X) & \Rightarrow & \phantom{\neg} \mi{fly}(X) \\
r_{2}: & \mi{penguin}(X) & \Rightarrow & \neg \mi{fly}(X) \\
r_{3}: & \mi{penguin}(X) & \rightarrow & \phantom{\neg} \mi{bird}(X) \\
r_{4}: & \mi{injured}(X)   & \leadsto     &  \neg \mi{fly}(X) \\
f      : & \mi{penguin}(\mi{tweety}) &  &  \\
g      : & \mi{bird}(\mi{freddie}) &  &  \\
h     : & \mi{injured}(\mi{freddie}) &  &  \\
\end{array}
\]
and a priority relation $r_{2} > r_{1}$.

Here $r_1, r_2, r_3, r_4, f, g, h$ are labels and
$r_3$ is (a reference to) a strict rule, while $r_1$ and $r_2$ are defeasible rules,
$r_4$ is a defeater,
and $f, g, h$ are facts.
Thus $F = \{f,g,h\}$, $R_s = \{ r_3 \}$, $R_{sd} =  \{r_1, r_2, r_3 \}$ and $R = \{r_1, r_2, r_3, r_4 \}$
and $>$ consists of the single tuple $(r_2, r_1)$.
The rules express that birds usually fly ($r_1$),
penguins usually don’t fly ($r_2$),
that all penguins are birds ($r_3$),
and that an injured animal may not be able to fly ($r_4$).
In addition, the priority of $r_{2}$ over $r_{1}$ expresses that when something is both a bird and a penguin (that is, when both rules can fire) it usually cannot fly
(that is, only $r_{2}$ may fire, it overrules $r_{1}$).
Finally, we are given the facts that $\mi{tweety}$ is a penguin,
and $\mi{freddie}$ is an injured bird.

This defeasible theory is hierarchical.
One function that demonstrates this
maps $\mi{injured}$ and $\mi{penguin}$ to 0,
$\mi{bird}$ to 1, and $\mi{fly}$ to 2.
\end{example}

A \emph{conclusion} takes the forms $+d \: q$ or $-d \: q$, where $q$ is a literal and $d$ is a tag indicating which inference rules were used.
Given a defeasible theory $D$,
$+d \: q$ expresses that $q$ can be proved via inference rule $d$ from $D$,
while $-d \: q$ expresses that it can be established that $q$ cannot be proved from $D$.

For example, in \cite{TOCL01}, a defeasible logic, now called $\DL(\partial)$, is defined
with the following inference rules, phrased as conditions on proofs\footnote{
Here, 
$D$ is a defeasible theory  $(F,R,>)$,
$q$ is a variable-free literal,
$P$ denotes a proof (a sequence of conclusions constructed by the inference rules), 
$P[1..i]$ denotes the first $i$ elements of $P$,
and $P(i)$ denotes the $i^{th}$ element of $P$.
}

\noindent\begin{minipage}[t]{.45\textwidth}
\begin{tabbing}
90123456\=7890\=1234\=5678\=9012\=3456\=\kill

$+\Delta)$  If  $P(i+1) = +\Delta q$  then either \\
\hspace{0.2in}  (1)  $q \in F$;  or \\
\hspace{0.2in}  (2)  $\exists r \in R_{s}[q] \  \forall a \in A(r),
+\Delta a \in P[1..i]$.
\end{tabbing}
\end{minipage}
\begin{minipage}[t]{.45\textwidth}
\begin{tabbing}
90123456\=7890\=1234\=5678\=9012\=3456\=\kill

$-\Delta)$  If  $P(i+1) = -\Delta q$  then \\
\hspace{0.2in}  (1)  $q \notin F$,  and \\
\hspace{0.2in}  (2)  $\forall r \in R_{s}[q] \  \exists a \in A(r),
-\Delta a \in P[1..i]. $
\end{tabbing}
\end{minipage}
\smallskip
\smallskip

These two inference rules concern reasoning about definitive information,
involving only strict rules and facts.
They define conventional monotonic inference ($+\Delta$)
and provable inability to prove from strict rules and facts ($-\Delta$).
For example, $+\Delta$ says that $+\Delta q$ can be added to a proof $P$ at position $i+1$ only if
$q \in F$  or
there is a strict rule $r$ with head $q$ where
each literal $a$ in the antecedent $A(r)$ has been proved ($+\Delta a$) earlier in the proof ($P[1..i]$).

The next rules refer to defeasible reasoning.

\smallskip
\smallskip
\noindent\begin{minipage}[t]{.45\textwidth}
\begin{tabbing}
$+\partial)$  If  $P(i+1) = +\partial q$  then either \\
\hspace{0.2in}  (1)  $+\Delta q \in P[1..i]$; or  \\
\hspace{0.2in}  (2)  The following three conditions all hold. \\
\hspace{0.4in}      (2.1)  $\exists r \in R_{sd}[q] \  \forall a \in A(r)$,  \\
\hspace{1.1in}                                  $+\partial a \in P[1..i]$,  and \\
\hspace{0.4in}      (2.2)  $-\Delta \non q \in P[1..i]$,  and \\
\hspace{0.4in}      (2.3)  $\forall s \in R[\non q]$  either \\
\hspace{0.6in}         (2.3.1)  $\exists a \in A(s),  -\partial a \in P[1..i]$;  or \\
\hspace{0.6in}          (2.3.2)  $\exists t \in R_{sd}[q]$  such that \\
\hspace{0.8in}                $\forall a \in A(t),  +\partial a \in P[1..i]$,  and \\
\hspace{0.8in}                $t > s$.
\end{tabbing}
\end{minipage}
\begin{minipage}[t]{.45\textwidth}
\begin{tabbing}
$-\partial)$  If  $P(i+1) = -\partial q$  then \\
\hspace{0.2in}  (1)  $-\Delta q \in P[1..i]$, and \\
\hspace{0.2in}  (2)  either \\
\hspace{0.4in}      (2.1)  $\forall r \in R_{sd}[q] \  \exists a \in A(r)$,  \\
\hspace{1.1in}                                        $-\partial a \in P[1..i]$; or \\
\hspace{0.4in}      (2.2)  $+\Delta \non q \in P[1..i]$; or \\
\hspace{0.4in}      (2.3)  $\exists s \in R[\non q]$  such that \\
\hspace{0.6in}          (2.3.1)  $\forall a \in A(s),  +\partial a \in P[1..i]$,  and \\
\hspace{0.6in}          (2.3.2)  $\forall t \in R_{sd}[q]$  either \\
\hspace{0.8in}                $\exists a \in A(t),  -\partial a \in P[1..i]$;  or \\
\hspace{0.8in}                not$(t > s)$.\\
\end{tabbing}
\end{minipage}

$+\partial q$ is a \emph{consequence} of a defeasible theory $D$ if there is a proof containing $+\partial q$.

In the $+\partial$ inference rule,
(1) ensures that any monotonic consequence is also a defeasible consequence.
(2) allows the application of a rule (2.1) with head $q$, provided that
monotonic inference provably cannot prove $\non q$ (2.2)
and every competing rule either provably fails to apply (2.3.1)
or is overridden by an applicable rule for $q$ (2.3.2).
The $-\partial$ inference rule is the strong negation \cite{flexf} of the $+\partial$ inference rule.
For other properties of this and other defeasible logics,
the reader is referred to \cite{TOCL10}.

\begin{example}    \label{ex:tweety2}
The above inference rules make several inferences
from the Tweety defeasible theory in Example \ref{ex:tweety}. 

The $+\Delta$ inference rule infers
$+\Delta\: \mi{penguin}(\mi{tweety})$,
$+\Delta\: \mi{bird}(\mi{freddie})$, 
$+\Delta\: \mi{injured}(\mi{freddie})$ 
from the facts, and
$+\Delta\: \mi{bird}(\mi{tweety})$
using $r_3$.
Such inferences are definite conclusions from the theory.
The $-\Delta$ inference rule infers, among others
$-\Delta\:  \mi{penguin}(\mi{freddie})$,
$-\Delta\: \mi{injured}(\mi{tweety})$,
$-\Delta\: \neg \mi{injured}(\mi{tweety})$, and
$-\Delta\: \neg \mi{bird}(\mi{tweety})$,
indicating that the theory is provably unable to come to a definite conclusion about these statements,
because there is no rule (and no fact) for these literals.
It also infers
$-\Delta\:  \mi{fly}(\mi{freddie})$,
$-\Delta\:  \neg \mi{fly}(\mi{freddie})$,
$-\Delta\:  \mi{fly}(\mi{tweety})$, and \linebreak
$-\Delta\: \neg  \mi{fly}(\mi{tweety})$
because there is no strict rule for $\mi{fly}$ or $\neg \mi{fly}$
and consequently (2) of the $-\Delta$ inference rule is vacuously true.

The $+\partial$ inference rule infers
$+\partial\: \mi{penguin}(\mi{tweety})$,
$+\partial\: \mi{bird}(\mi{freddie})$, and
$+\partial\: \mi{injured}(\mi{freddie})$, and 
$+\partial\: \mi{bird}(\mi{tweety})$
because these statements are known definitely.
It also concludes \linebreak
$+\partial\:  \neg  \mi{fly}(\mi{tweety})$
using rule $r_2$ in (2.1),
the previous conclusion 
$-\Delta\:  \mi{fly}(\mi{tweety})$ in (2.2),
and, despite the presence of $r_1$ as $s$ in (2.3),
using $r_2$ as $t$ in (2.3.2) with the priority statement $r_{2} > r_{1}$
to overrule $r_1$.
It is unable to similarly conclude 
$+\partial\:  \mi{fly}(\mi{freddie})$,
because of the presence of $r_3$ and the lack of a priority statement to overrule it.

The $-\partial$ inference rule infers, among others,
$-\partial\:  \mi{penguin}(\mi{freddie})$ and
$-\partial\: \mi{injured}(\mi{tweety})$
because these statements are known unprovable definitely (1),
and (2.1) is satisfied vacuously because there is no rule for these predicates.
It also infers
$-\partial\:  \mi{fly}(\mi{freddie})$,
$-\partial\:  \neg \mi{fly}(\mi{freddie})$, and
$-\partial\:  \mi{fly}(\mi{tweety})$.
\end{example}

\vspace{0.2cm}

For clarity, we refer to elements of defeasible theories as rules,
and elements of logic programs as clauses.
Also, note the distinction between rules (syntactic elements of a defeasible theory)
and inference rules (criteria for extending proofs).
Logic programming predicates will be written in $\mt{teletype}$ font,
while defeasible logic predicates will be written in $\mi{italics}$.
We use $not$ and $\mt{not}$ for negation-as-failure in logic programs,
and $\neg$ for classical negation in defeasible theories.
However, ``predicate'', ``atom'' and ``literal'' may be used to refer to elements
of either a defeasible theory or a logic program.

To avoid the confusion of  existing names with names generated during transformations,
we need a character that is not used in $D$.
For readability in this paper, we choose the underscore $\_$ as this character,
but any other character would suffice.

\section{The scalable defeasible logic $\DL(\pl)$}\label{sec:sdl}

The defeasible logic  $\DL(\pl)$  \cite{sdl} was designed to allow defeasible inference
to be scalable to very large data sets.
We present the inference rules of that logic here.

$\DL(\pl)$ involves three tags: 
$\Delta$, which expresses conventional monotonic inference;
$\lambda$, an auxiliary tag;
and $\pl$, which is the main notion of defeasible proof in this logic.
The inference rules are presented below, phrased as conditions on proofs\footnote{
As in the inference rules for $\DL(\partial)$ in the previous section,
$D$ is a defeasible theory  $(F,R,>)$,
$q$ is a variable-free literal,
$P$ denotes a proof, 
$P[1..i]$ denotes the first $i$ elements of $P$,
and $P(i)$ denotes the $i^{th}$ element of $P$.
}.

\begin{tabbing}
90123456\=7890\=1234\=5678\=9012\=3456\=\kill

$+\Delta)$  If  $P(\imath+1) = +\Delta q$  then either \\
\hspace{0.2in}  (1)  $q \in F$;  or \\
\hspace{0.2in}  (2)  $\exists r \in R_{s}[q] \  \forall a \in A(r),
+\Delta a \in P[1..\imath]$.
\end{tabbing}
\smallskip

This inference rule concerns reasoning about definitive information,
involving only strict rules and facts.
It is identical to the rule for monotonic inference in $\DL(\partial)$.

For a defeasible theory $D$,
we define $P_{\Delta}$ to be 
the set of consequences in the largest proof satisfying the proof condition $+\Delta$,
and call this the $\Delta$ \emph{closure}.
It contains all $+\Delta$ consequences of $D$.

Once $P_{\Delta}$ is computed, we can apply the $+\lambda$ inference rule.
$+\lambda q$ is intended to mean that $q$ is potentially defeasibly provable in $D$.
The $+\lambda$ inference rule is as follows.

\begin{tabbing}
$+\lambda$: \= If $P(\imath + 1) = +\lambda q$ then either \\
\> (1) \=$+\Delta q \in P_{\Delta}$ or \\
\> (2)	\>(2.1) $\exists r \in R_{sd}[q] ~ \forall \alpha \in A(r): +\lambda \alpha \in P(1..\imath)$ and \\
\> \>(2.2) $+\Delta \non q \notin P_{\Delta}$ 
\end{tabbing}
\smallskip

Using this inference rule, and given $P_{\Delta}$, we can compute the $\lambda$ closure $P_{\lambda}$,
which contains all $+\lambda$ consequences of $D$.

$+\pl q$ is intended to mean that $q$ is defeasibly provable in $D$.
Once $P_{\Delta}$ and $P_{\lambda}$ are computed, we can apply the $+\pl$ inference rule.

\begin{tabbing}
$+\pl$: \= If $P(\imath + 1) = +\pl q$ then either \\
\> (1) \=$+\Delta q  \in P_{\Delta}$ or \\
\> (2)	\>(2.1) $\exists r \in R_{sd}[q] ~ \forall \alpha \in A(r): +\pl \alpha \in P(1..\imath)$ and \\
\> \>(2.2) $+\Delta \non q \notin P_{\Delta}$ and \\
\> \>(2.3) \=$\forall s \in R[\non q]$ either \\
\> \> \>(2.3.1) $\exists \alpha \in A(s): +\lambda \alpha \notin P_{\lambda}$ or \\
\> \>\>(2.3.\=2) $\exists t \in R_{sd}[q]$ such that \\
\> \>\>\>$\forall \alpha \in A(t): +\pl \alpha \in P(1..\imath)$ and $t > s$
\end{tabbing}

The $\pl$ closure $P_{\pl}$ contains all $\pl$ consequences of $D$.
We say a set of tagged literals $Q$ is \emph{$\pl$-deductively closed} if,
given closures $P_\Delta$ and $P_\lambda$ as defined above,
 for every literal $q$ that
may be appended to $Q \cup P_\Delta \cup P_\lambda$ by the inference rule $+\pl$ (treating $Q \cup P_\Delta \cup P_\lambda$ as a proof), $q \in Q$.
Clearly, $P_{\pl}$ is the smallest $\pl$-deductively closed set.

\begin{example}    \label{ex:tweety3}
We now apply these
inference rules to the Tweety defeasible theory in Example \ref{ex:tweety}. 

As before, the $+\Delta$ inference rule infers
$+\Delta\: \mi{penguin}(\mi{tweety})$,
$+\Delta\: \mi{bird}(\mi{freddie})$, and
\linebreak
$+\Delta\: \mi{injured}(\mi{freddie})$ 
from the facts, and
$+\Delta\: \mi{bird}(\mi{tweety})$
using $r_3$.
We have no need of the $-\Delta$ inference rule.

Using the $+\lambda$ inference rule
we infer all the literals inferred by the $+\Delta$ inference rule as $+\lambda$ conclusions.
In addition, the rule infers
$+\lambda\: \mi{fly}(\mi{tweety})$,
$+\lambda\: \mi{fly}(\mi{freddie})$, and
$+\lambda\: \neg\mi{fly}(\mi{tweety})$.

Using the $+\pl$ inference rule,
again all the $+\Delta$ conclusions are inferred as $+\pl$ conclusions.
The only other conclusion that can be drawn with this rule is
$+\pl\: \neg\mi{fly}(\mi{tweety})$.
The potential inference of $\mi{fly}(\mi{tweety})$ is overruled by $r_2$ inferring $+\pl\: \neg\mi{fly}(\mi{tweety})$.
On the other hand, a potential inference of $\mi{fly}(\mi{freddie})$ is not obtained
because $r_1$ cannot overrule $r_4$.
\end{example}

Inference rules $\partial$ and $\pl$ employ the notion of ``team defeat'',
where it doesn't matter which rule overrides an opposing rule, as long as all opposing rules are overridden.
This is expressed in (2.3.2).
We can also have a version of $\pl$ with ``individual defeat'',
where all opposing rules must be overridden by the same rule,
which we denote by $\pl^*$.
The inference rule for $+\pl^*$ replaces (2.3.2) in $\pl$ by $r > s$.

\begin{tabbing}
$+\pl^*$: \= If $P(\imath + 1) = +\pl^* q$ then either \\
\> (1) \=$+\Delta q  \in P_{\Delta}$ or \\
\> (2)	\>(2.1) $\exists r \in R_{sd}[q] ~ \forall \alpha \in A(r): +\pl^* \alpha \in P(1..\imath)$ and \\
\> \>(2.2) $+\Delta \non q \notin P_{\Delta}$ and \\
\> \>(2.3) \=$\forall s \in R[\non q]$ either \\
\> \> \>(2.3.1) $\exists \alpha \in A(s): +\lambda \alpha \notin P_{\lambda}$ or \\
\> \>\>(2.3.\=2)  $r > s$
\end{tabbing}

Example~\ref{ex:tweety3} does not display the distinction between $\pl$ and $\pl^*$:
both have the same consequences.
The distinction is visible when there are multiple applicable rules for both some literal $q$ and its negation $\non q$.
The following example originates from \cite{TOCL01}.

\begin{example}  \label{ex:indiv}
Consider some rules of thumb about animals and, particularly, mammals.
An egg-laying animal is generally not a mammal.
Similarly, an animal with webbed feet is generally not a mammal.
On the other hand, an animal with fur is generally a mammal.
Finally, the monotremes are a subclass of mammal.
These rules are represented as defeasible rules below.

Furthermore, animals with fur and webbed feet are generally mammals,
so $r_2$ should overrule $r_4$.
And monotremes are a class of egg-laying mammals,
so $r_1$ should overrule $r_3$.

Finally, it happens that a platypus is
a furry, egg-laying, web-footed monotreme.
Is it a mammal?
(That is, is $mammal(platypus)$ a consequence of the defeasible theory below?)
\begin{tabbing}

123412341234123412\=34123412\=3412\=34123\=41234123412341234\=\kill

$r_1: \ monotreme(X)$ \> $\Rightarrow mammal(X)$
\>\>\> $r_3: \ laysEggs(X)$ \> $\Rightarrow \neg mammal(X)$ \\

$r_2: \ hasFur(X)$ \> $\Rightarrow mammal(X)$
\>\>\> $r_4: \ webFooted(X)$ \> $\Rightarrow \neg mammal(X)$ \\

$r_1 > r_3$ \>\>\>\>  $r_2 > r_4$ \\

$monotreme(platypus)$ \>\>\>\> $laysEggs(platypus)$ \\

$hasFur(platypus)$ \>\>\>\> $webFooted(platypus)$ \\

\end{tabbing}

It is obvious that all four rules are applicable to the question of $mammal(platypus)$.
Under team defeat, each rule for $\neg mammal(platypus)$ is overcome by some rule for $mammal(platypus)$,
so $mammal(platypus)$ is inferred (which is zoologically correct).
However, there is no single rule for $mammal(platypus)$ that overcomes all rules for $mammal(platypus)$,
so under individual defeat we cannot infer $mammal(platypus)$ (nor $\neg mammal(platypus)$).
\end{example}

This logics are amenable to the techniques used to establish
model-theoretic \cite{Maher02} and argumentation \cite{JLC04} semantics
for other defeasible logics.
However, the proof-based definitions defined above are sufficient for this paper.

A key feature of $\DL(\pl)$ and $\DL(\pl^*)$ inference rules is that
they do not use negative inference rules, unlike  $\DL(\partial)$ which uses $-\Delta$ and $-\partial$.
Instead they use expressions $+\Delta q \notin P_\Delta$ and $+\lambda q \notin P_\lambda$.
This choice was founded on practical difficulties in 
scalably implementing existing non-propositional defeasible logics \cite{sdl,ECAI2012}.
However, it has implications for the structure and semantics of the corresponding metaprograms
of the logics, as discussed in the next section.

\section{Metaprogram for $\DL(\pl)$}  \label{sec:metaprogram}

Following \cite{MG99,flexf}, we can map $\DL(\pl)$ to a logic program
by expressing the inference rules of $\DL(\pl)$ as a metaprogram.
The metaprogram assumes that a defeasible theory $D=(F,R,>)$ is represented as a set of unit clauses, as follows.

\begin{enumerate}
\item $\mt{fact}(p)$.  \hfill if $p\in F$
\item $\mt{strict}(r,p,[q_1,\dots,q_n])$. 
\hspace*{\fill}
if
  $r:q_1,\dots,q_n\to p\in R$
\item $\mt{defeasible}(r,p,[q_1,\dots,q_n])$.
\hspace*{\fill}
if
  $r:q_1,\dots,q_n\Rightarrow p\in R$
\item $\mt{defeater}(r,p,[q_1,\dots,q_n])$. 
\hspace*{\fill}
if 
  $r:q_1,\dots,q_n\leadsto p\in R$
\item $\mt{sup}(r,s)$. 
\hspace*{\fill}
for each pair of rules such that
  $r > s$
\end{enumerate}

Note that predicates in $D$ are represented as function symbols.
For convenience, we assume that negated atoms $\neg p(\vec{a})$ in the defeasible theory are represented by $not\_p(\vec{a})$.

As in \cite{MG99,flexf}, the metaprogram is presented for ease of understanding,
rather than formal logic programming syntax.
The full details are available in the appendix.

\begin{Clause}\label{strictly1}
  $\mt{definitely}(X)$\ :-\\
  \> $\mt{fact}(X)$.
\end{Clause}

\begin{Clause}\label{strictly2}
  $\mt{definitely}(X)$\ :-\\
  \> $\mt{strict}(R,X,[\seq{Y}])$,\\
  \> $\mt{definitely}(Y_1)$, \dots, $\mt{definitely}(Y_n)$.
\end{Clause}

\begin{Clause}\label{lambda1}
  $\mt{lambda}(X)$\ :-\\
  \> $\mt{definitely}(X)$.
\end{Clause}

\begin{Clause}\label{lambda2}
  $\mt{lambda}(X)$\ :-\\
  \> $\mt{not\ definitely}(\non X)$,\\
  \> $\mt{strict\_or\_defeasible}(R,X,[\seq{Y}])$,\\
  \> $\mt{lambda}(Y_1)$, \dots, $\mt{lambda}(Y_n)$.
\end{Clause}

\begin{Clause}\label{defeasibly1}
  $\mt{defeasibly}(X)$\ :-\\
  \> $\mt{definitely}(X)$.
\end{Clause}

\begin{Clause}\label{defeasibly2}
  $\mt{defeasibly}(X)$\ :-\\
  \> $\mt{not\ definitely}(\non X)$,\\
  \> $\mt{strict\_or\_defeasible}(R,X,[\seq{Y}])$,\\
  \> $\mt{defeasibly}(Y_1)$, \dots, $\mt{defeasibly}(Y_n)$,\\
  \> $\mt{not\ overruled}(X)$.
\end{Clause}

\begin{Clause}\label{overruled}
  $\mt{overruled}(X)$\ :-\\
  \> $\mt{rule}(S,\non X,[\seq{U}])$,\\
  \> $\mt{lambda}(U_1)$, \dots, $\mt{lambda}(U_n)$,\\
  \> $\mt{not\ defeated}(S,\non X)$.
\end{Clause}

\begin{Clause}\label{defeated}
  $\mt{defeated}(S,\non X)$\ :-\\
  \> $\mt{sup}(T,S)$, \\
  \> $\mt{strict\_or\_defeasible}(T,X,[\seq{V}])$,\\
  \> $\mt{defeasibly}(V_1)$, \dots, $\mt{defeasibly}(V_n)$.
\end{Clause}

For $\DL(\pl^*)$, clauses~\ref{defeasibly2}, \ref{overruled} and \ref{defeated} are replaced by the following.
(Clauses~\ref{defeasibly2} and \ref{overruled} are modified only slightly.)

\begin{Clause}\label{defeasibly2_indiv}
  $\mt{defeasibly}(X)$:-\\
  \> $\mt{not\ definitely}(\non X)$,\\
  \> $\mt{strict\_or\_defeasible}(R,X,[\seq{Y}])$,\\
  \> $\mt{defeasibly}(Y_1)$, \dots, $\mt{defeasibly}(Y_n)$,\\
  \> $\mt{not\ overruled}(R,X)$.
\end{Clause}

\begin{Clause}\label{overruled_indiv}
  $\mt{overruled}(R,X)$:-\\
  \> $\mt{rule}(S,\non X,[\seq{U}])$,\\
  \> $\mt{lambda}(U_1)$, \dots, $\mt{lambda}(U_n)$,\\
  \> $\mt{not\ defeats}(R,S)$.
\end{Clause}

\begin{Clause}\label{defeated_indiv}
  $\mt{defeats}(R,S)$:-\\
  \> $\mt{sup}(R,S)$. 
\end{Clause}

We use $\M_{\pl}$ to denote the metaprogram for $\DL(\pl)$ 
(clauses \ref{strictly1}--\ref{defeated}) %and \ref{conc_definitely}--\ref{conc_defeasibly})
and $\M_{\pl^*}$  to denote the metaprogram for $\DL(\pl^*)$
(clauses \ref{strictly1}--\ref{defeasibly1}  and \ref{defeasibly2_indiv}--\ref{defeated_indiv}), 
% and \ref{conc_definitely}--\ref{conc_defeasibly})
or simply $\M$ to refer to either metaprogram.
The combination of $\M$ and the representation of $D$ is denoted by $\M(D)$.

We interpret $\M(D)$ under the well-founded semantics \cite{WF91}.
Notice that \ref{strictly1} -- \ref{lambda2}, together with the representation of $D$,
form a stratified logic program, with $\mt{definitely}$ and $\mt{lambda}$ in different strata.
The choice of well-founded semantics means that the explicit ordering of computing $P_\Delta$, then $P_\lambda$,
then $P_{\pl}$ is implemented implicitly by the stratification.
For weaker semantics that are not equal to the stratified semantics on stratified programs,
like Fitting's semantics, the ordering would need to be expressed explicitly.

A careful comparison of the parts of the inference rules $+\Delta$, $+\lambda$, $+\pl$, and $+\pl^*$
with the clauses of the metaprogram strongly suggests the correctness of the metaprogram
representation.
Clauses \ref{strictly1} and \ref{strictly2} represent the $+\Delta$ inference rule,
and clauses  \ref{lambda1} and  \ref{lambda2} represent the $+\lambda$ inference rule.
Clauses \ref{defeasibly1}--\ref{defeated} represent the $+\pl$ inference rule:
clause  \ref{defeasibly1} corresponds to (1) of the inference rule; 
\ref{defeasibly1} corresponds to (2), with the body expressing (2.1) and (2.2) and
the negated call to $\mt{overruled}$ representing (2.3);
\ref{overruled} corresponds to (2.3.1); and
\ref{defeated} corresponds to (2.3.2).
Negations are used to express the universal quantifier in (2.3) via the implicit logic programming quantifier
and $\forall = \neg \exists \neg$.
However, despite this close correspondence,
we still need to establish correctness formally, and that proof will be easier using a transformed program. 
Consequently, the proof that $\M(D)$ under the well-founded semantics correctly represents the logic defined in Section \ref{sec:sdl} is deferred to Section \ref{sec:semantics}.

\vspace{0.2cm}

There have been several mappings of defeasible logics to logic programs.
An early implementation of a defeasible logic, d-Prolog \cite{d-Prolog,Nute_subsumption},
was defined as a Prolog metaprogram.
Courteous Logic Programs \cite{Grosof97} were originally implemented by directly modifying rules to express overriding,
% IBM Research Report RC 21472
but later versions (such as \cite{LPDA}) used a metaprogramming approach.
Inspired by d-Prolog, \cite{MG99} defined the metaprogram approach for $\DL(\partial)$, 
mapping defeasible theories to logic programs under Kunen's semantics.  
(See also \cite{TPLP06}.)
A key point of \cite{MG99} was the decomposition of defeasible logics into
a conflict resolution method, expressed by the metaprogram,
and a notion of failure, expressed by the logic programming semantics applied to the metaprogram.
That paper also introduced well-founded defeasible logic ($WFDL$) and mapped it to
logic programs under the well-founded semantics, as an example of this decomposition. 
\cite{flexf} extended that approach to other defeasible logics,
and developed a principled framework for defeasible logics.
\cite{GM17} further extended this work to a single formalism supporting multiple methods of
conflict resolution.

\cite{embed,TPLP06} investigated mappings of defeasible theories in $\DL(\partial)$ to logic programs under the stable semantics, but achieved only partial results. 
These works used the metaprogram as a basis,
but \cite{embed} provided a simpler mapping of propositional defeasible theories in $\DL(\partial)$ to logic programs,
under the assumption that $D$ was simplified by the transformations of \cite{TOCL01}.
\cite{Brewka01} mapped the ambiguity propagating defeasible logic $\DL(\delta)$
to prioritized logic programs under the well-founded semantics \cite{Brewka96},
and showed that the mapping is not sound, nor complete.
Brewka concluded that differences in the treatment of strict rules affected completeness,
while differences in semantics (well-founded versus Kunen)
and in treatment of rule priorities affected  soundness.
From the viewpoint of \cite{MG99},
these works demonstrate that a structured mapping of a defeasible logic under one semantics to logic programming under a different semantics is difficult.
(See also \cite{Maher13b}.)

More  recently, \cite{MN06,MN10,Maier13} 
provided mappings, based on earlier mappings, 
of the defeasible logic ADL to logic programs under the well-founded semantics, and another mapping in the reverse direction.
ADL is similar to the well-founded defeasible logic $\WFDL(\delta^*)$ \cite{Maher14},
but with a more sophisticated treatment of inconsistency of literals.
\cite{Maier13} also discusses a stable set semantics for these logics,
and relates it to logic programs under the stable semantics.
% \finish{Tommie Meyer paper?? \cite{DDLV} NO - it is an ad hoc generation of priorities with intuition similar
% to Nute's subsumption-based priorities.  There is no independent semantics.}

These works all focussed on the correctness of the mapping.
Beyond d-Prolog and Courteous Logic Programs, there is little attention to implementation, and
there is no analysis of the structure of the resulting logic programs.
That may be because there is little useful structure to those programs.
In contrast, both $\M_{\pl}$ and $\M_{\pl^*}$ have a convenient structure.\footnote{
Recall that $\M_{\pl}$ is defined in  \ref{app:metaprogram}, with a more readable version expressed above.
The following propositions and discussion refer to the syntax in the appendix.
}

\begin{propn}   \label{prop:stratified}
For any defeasible theory $D$,
$\M_{\pl}(D)$ is call-consistent and
$\M_{\pl^*}(D)$ is stratified.
\end{propn}
\skipit{
\begin{proof}
Only the predicates 
$\mt{defeasibly}$, $\mt{overruled}$, $\mt{defeated}$, and $ \mt{loop\_defeasibly}$
in $\M_{\pl}$
are related by $\approx$.
From the clauses we only have
$\mt{defeasibly} \mdn \mt{overruled}$,
$\mt{overruled} \mdn \mt{defeated}$, 
$\mt{defeasibly} \mdp \mt{loop\_defeasibly}$,
$\mt{defeated} \mdp \mt{loop\_defeasibly}$, and
and $ \mt{loop\_defeasibly}  \mdp \mt{defeasibly}$.
Thus, for none of these predicates (or any others) do we have $p \gen p$.
Consequently $\M_{\pl}$ is call-consistent.

Consider the following ordering on predicates in $\M_{\pl^*}(D)$.
\[
\begin{array}{c}
\mt{defeasibly}, \mt{loop\_defeasibly} \\
\uplt \\
\mt{lambda},  \mt{loop\_lambda}, \mt{overruled} \\ 
\uplt \\
\mt{definitely}, \mt{loop\_definitely}, \mt{defeats} \\
\uplt \\
\mt{neg}, \mt{sup}, \mt{fact}, \mt{strict}, \mt{defeasible}, \mt{defeater} ,
\mt{strict\_or\_defeasible}, \mt{rule}  
\end{array}
\]
It is straightforward to verify that this provides a stratification of $\M_{\pl^*}(D)$.
\end{proof}
}
In contrast, the metaprograms for $\DL(\partial)$ and $\DL(\partial^*)$ \cite{flexf} are not call-consistent,
resulting from the use of $-\partial$ ($-\partial^*$) in (2.3.1).

The metaprogram $\M_{\pl}$ is not strict:
both $\mt{defeasibly}$ and $\mt{lambda}$ depend on $\mt{definitely}$
both positively and negatively.
Furthermore, $\mt{defeasibly}$ depends on $\mt{strict\_or\_defeasible}$ and $\mt{neg}$
both positively (via $\mt{defeated}$)
and negatively (via $\mt{lambda}$).
Thus $\M_{\pl}$ does not have a signing.
However, if we take a floor consisting of $\mt{definitely}$, $\mt{lambda}$ and all supporting predicates
then the remainder of $\M_{\pl}$ does have a signing.

\begin{propn}
Let $D$ be a defeasible theory, and
let $\Q = \{ \mt{defeasibly},  \mt{overruled}, \mt{defeated}, \mt{loop\_defeasibly} \}$.
\begin{itemize}
\item
$\M_{\pl}(D)$ has a signing $s$ for $\Q$
such that 
$\mt{defeasibly}$ has sign $+1$ and avoids negative unfoundedness wrt $s$ and $\Q$.
\item
$\M_{\pl^*}(D)$ has a signing $s$ for $\Q$
such that 
$\mt{defeasibly}$ has sign $+1$ and avoids negative unfoundedness wrt $s$ and $\Q$.
\end{itemize}
\end{propn}
\skipit{
\begin{proof}
Let $\P$ be the set of predicates occurring in $\M_{\pl}$ and 
let $\F$ be $\P \backslash \Q$.
%$\F = \{ \mt{definitely}, \mt{lambda} \}$.
Then $\P$ is a downward-closed set of predicates with floor $\F$.
Furthermore, $\Q$ has a signing $s$ that maps
$\mt{defeasibly}$, $\mt{defeated}$ and $\mt{loop\_defeasibly}$ to $+1$
and $\mt{overruled}$ to $-1$.
$\mt{defeasibly}$ avoids negative unfoundedness wrt $s$ and $\Q$,
because the only negatively signed predicate is $\mt{overruled}$,
and $\mt{overruled}$ does not depend positively on any predicate in $\Q$.

The same reasoning applies for $\M_{\pl^*}$ with the same signing $s$.
 \end{proof}
}

Notice that we can get a signing for $\M_{\pl}$ and $\M_{\pl^*}$ with smaller floor 
(by excluding $\mt{lambda}$ and  $\mt{loop\_lambda}$),
but then $\mt{defeasibly}$ does not necessarily avoid negative unfoundedness.
For example, if $D$ consists of $p \Rightarrow p$ then $\mt{lambda}$, which would have a negative sign,
has an unfounded sequence $\mt{lambda}(p), \mt{loop\_lambda}([p]), \mt{lambda}(p), \ldots$.

As a consequence of the previous proposition, Theorem \ref{thm:comb} part \ref{thm:EQ} applies.
This means that, once the values of the predicates in $\F$ (particularly $\mt{definitely}$ and $\mt{lambda}$)
are determined,
the true $\mt{defeasibly}$ atoms can be computed under either the Fitting or well-founded semantics.
This reflects the original definition of $\DL(\pl)$ in Section \ref{sec:sdl},
where $P_\Delta$ and $P_\lambda$ must be computed before computing $P_{\pl}$.

Similarly, by Theorem ~\ref{thm:comb} part \ref{thm:new}, if computing using the well-founded semantics,
only the positive parts of $\mt{defeasibly}$, $\mt{defeated}$ and $\mt{loop\_defeasibly}$ 
need be computed, and only the negative part of $\mt{overruled}$ is needed.
In contrast, in the corresponding metaprogram $\M_{\partial}$ for $\DL(\partial)$
(see \cite{MG99,flexf})
$\mt{defeasibly}$ depends negatively on itself, via $\mt{overruled}$.
Consequently,  $\M_{\partial}$ is not call-consistent and there is no useful signing for $\M_{\partial}$
\footnote{We could set $\Q = \{ \mt{defeasibly} \}$, to get a signing, but then $\P\backslash\Q$
is not downward-closed
(because $\mt{overruled}$ depends on $\mt{defeasibly}$), 
and so cannot serve as a floor.
Thus, Theorem ~\ref{thm:comb} cannot be applied.},
nor for $\M_{\partial^*}$.
Thus, we see that the structure of $\DL(\pl)$ (and, hence, of $\M_{\pl}$) allows optimizations,
like those of Theorem~\ref{thm:comb},
that are unavailable to $\M_{\partial}$ and $\M_{\partial^*}$.

Not all structural aspects of the metaprogram are convenient.
$\M_{\pl}$ is safe, but the representation of $D$ may not be.
Furthermore, $\M_{\pl^*}$ is not safe.
In fact,
clause \ref{overruled_indiv} is neither range-restricted nor negation-safe,
because the variable $R$ appears in the head and negative literal, but not in a positive body literal.
Furthermore, even when $D$ is function-free, if it is not propositional then $\M_{\pl}(D)$ is not a \Datalog{}
program, because the predicates of $D$ are represented as functions in $\M_{\pl}(D)$.
These problems will be addressed by the transformation of $\M_{\pl}(D)$ in the next section.

\section{Transforming the Metaprogram}   \label{sec:trans}

We seek a logic program that is equivalent to $\M(D)$
on $\mt{definitely}$, $\mt{defeasibly} $ and $\mt{lambda}$ atoms,
but is in \Datalog{} form.
We manipulate $\M(D)$, using transformations that preserve the semantics of the program,
to achieve this end.
Specifically, we use a series of fold and unfold transformations \cite{TS,trans}.
These are known to preserve the well-founded semantics  \cite{unstable,Aravindan}.
We also introduce rules for new predicates, and delete rules for predicates that are no longer used.
These preserve the semantics of the important predicates \cite{unstable}.

\subsection{Partial Evaluation of the Metaprogram}

The following transformations are applied to the full metaprogram in  \ref{app:metaprogram}
and the representation of a theory $D$.
Unfold all occurrences of $\mt{rule}$, $\mt{strict\_or\_defeasible}$, $\mt{fact}$, $\mt{strict}$, $\mt{defeasible}$, and $\mt{defeater}$, to create specialized versions of the clauses for each applicable rule.
Unfold all occurrences of $\mt{neg}$, implementing $\non X$,
and unroll (i.e. repeatedly unfold) all occurrences of $\mt{loop\_definitely}$, $\mt{loop\_defeasibly}$ and $\mt{loop\_lambda}$,
implementing iteration over each body literal.
Because all lists are (terminated) finite lists, unrolling will terminate.

After this process, for any strict or defeasible rule, say
\[
\begin{array}{lrcl}
t: &   p(X, Z), \neg p(Z, Y)        & \Rightarrow & q(X, Y) \\
\end{array}
\]
we have a corresponding version of clause~\ref{defeated}
\begin{clause}
  $\mt{defeated}(S, not\_q(X, Y))$\ :-\\
  \> $\mt{sup}(t,S)$, \\
  \> $\mt{defeasibly}(p(X, Z))$, $\mt{defeasibly}(not\_p(Z, Y))$.
\end{clause}
and similar versions of clause~\ref{overruled}, etc.
Unfolding all occurrences of $\mt{sup}$ leaves us with clauses
\begin{clause}
  $\mt{defeated}(s, not\_q(X, Y))$\ :-\\
  \> $\mt{defeasibly}(p(X, Z))$, $\mt{defeasibly}(not\_p(Z, Y))$.
\end{clause}
for every rule $s$ with  $t > s$ in $D$.
Note that if $t$ is not superior to any rule then no clause is generated for $t$.

We then delete all clauses defining the predicates we have unfolded.  Such a deletion is correct because the three
predicates we are interested in no longer depend on the deleted predicates.

After all these transformations,
the only function symbols left are the predicates $p$, and their counterparts $not\_p$, that originate in the representation of $D$.
The resulting program $P_D$ is a partial evaluation of $\M_{\pl}$ wrt the representation of $D$.
It is a particularly transparent translation of the defeasible theory into a logic program.

\begin{example}
Consider a defeasible theory consisting of the rules
\[
\begin{array}{lrcl}
s: &   p(X, Y), q(Y, X)               & \Rightarrow & \neg q(X, Y) \\
t: &   p(X, Z), \neg p(Z, Y)        & \Rightarrow & \phantom{\neg} q(X, Y) \\
\end{array}
\]
with $t > s$.

The transformed program contains
\[
\begin{array}{lrcl}
  & \mt{defeasibly}(not\_q(X, Y))  & \mbox{:-} &  \\
 &&& \mt{not\ definitely}(q(X, Y)),  \\
 &&&  \mt{defeasibly}(p(X, Y)),  \mt{defeasibly}(q(Y, X)), \\
 &&& \mt{not\ overruled}(not\_q(X, Y)). \\
 
 &  \mt{overruled}(q(X, Y))  & \mbox{:-} &  \\
 &&& \mt{lambda}(p(X, Y)), \mt{lambda}(q(Y, X)), \\
 &&& \mt{not\ defeated}(s, not\_q(X, Y)). \\
 
& \mt{defeasibly}(q(X, Y))  & \mbox{:-} &  \\
 &&& \mt{not\ definitely}(not\_q(X, Y)),  \\
 &&&  \mt{defeasibly}(p(X, Z)),  \mt{defeasibly}(not\_p(Z, Y)), \\
 &&& \mt{not\ overruled}(q(X, Y)). \\
 
 &  \mt{overruled}(not\_q(X, Y))  & \mbox{:-} &  \\
 &&& \mt{lambda}(p(X, Z)), \mt{lambda}(not\_p(Z, Y)), \\
 &&& \mt{not\ defeated}(t, q(X, Y)). \\
 
 & \mt{defeated}(s, not\_q(X, Y)) & \mbox{:-} &  \\
 &&& \mt{defeasibly}(p(X, Z)), \mt{defeasibly}(not\_p(Z, Y)). \\
\end{array}
\]
as well as other clauses defining $\mt{lambda}$.
The first two clauses are derived from rule $s$ and the last three come from rule $t$.
There is no $ \mt{defeated}$ clause from $s$ because unfolding $\mt{sup}(s,S)$
eliminates the clause, there being no rule that $s$ is superior to.
\end{example}

The correctness of the modified program is straightforward.

\begin{propn}    \label{prop:trans1}
Let $D$ be a defeasible theory, and
let $P_D$ be the transformed version of $\M_{\pl}(D)$.
Let $q$ be a literal.
\begin{itemize}
\item
$\M_{\pl}(D) \models_{WF}  \mt{definitely}(q)$ iff $P_D \models_{WF}  \mt{definitely}(q)$
\item
$\M_{\pl}(D) \models_{WF}  \mt{lambda}(q)$ ~~~~~~~~\! iff $P_D \models_{WF}  \mt{lambda}(q)$
\item
$\M_{\pl}(D) \models_{WF}  \mt{defeasibly}(q)$ iff $P_D \models_{WF}  \mt{defeasibly}(q)$
\end{itemize}
\end{propn}
\skipit{
\begin{proof}
$\M(D)$ is transformed to $P_D$ by a series of unfolding and  deletion transformations
that preserve the well-founded semantics \cite{unstable,Aravindan}.
\end{proof}
}

\subsection{More transformation}

We now further transform $P_D$,
to make it more compact and to convert it to \Datalog{}.
To begin, we introduce new predicates and clauses.
For any literal $A$ (of the form $q(\vec{a})$ or $not\_q(\vec{a})$), we use $args(A)$ to denote $\vec{a}$.
For each rule
$
\begin{array}{lrcl}
r: &   B_1, \ldots, B_n       & \ARROW & A \\
\end{array}
$
in $D$, we add the clauses
\begin{Clause}  \label{body:def}
  $\mt{body}_r^d(args(A))$\ :-\\
  \> $\mt{defeasibly}(B_1)$, \dots, $\mt{defeasibly}(B_n)$.
\end{Clause}
\begin{Clause}   \label{body:lam}
  $\mt{body}_r^\lambda(args(A))$\ :-\\
  \> $\mt{lambda}(B_1)$, \dots, $\mt{lambda}(B_n)$.
\end{Clause}
If the rule is strict we also add
\begin{Clause}   \label{body:strict}
  $\mt{body}_r^\Delta(args(A))$\ :-\\
  \> $\mt{definitely}(B_1)$, \dots, $\mt{definitely}(B_n)$.
\end{Clause}

We then fold clauses derived from clauses \ref{defeasibly2} and \ref{defeated} by clause \ref{body:def},
fold clauses derived from clauses \ref{lambda2} and \ref{overruled} by clause \ref{body:lam}, and
fold clauses derived from clauses \ref{strictly2} by clause \ref{body:strict}.

This results in clauses of the form
\begin{clause}%\label{strictly2}
  $\mt{definitely}(A)$\ :-\\
  \> $\mt{body}_r^\Delta(args(A))$.
\end{clause}
\begin{clause}%\label{lambda2}
  $\mt{lambda}(A)$\ :-\\
  \> $\mt{not\ definitely}(\non A)$,\\
  \> $\mt{body}_r^\lambda(args(A))$.
\end{clause}
\begin{clause}%\label{defeasibly2}
  $\mt{defeasibly}(A)$\ :-\\
  \> $\mt{not\ definitely}(\non A)$,\\
  \> $\mt{body}_r^d(args(A))$,\\
  \> $\mt{not\ overruled}(A)$.
\end{clause}
\begin{clause}%\label{overruled}
  $\mt{overruled}(A)$\ :-\\
  \> $\mt{body}_s^\lambda(args(A))$,\\
  \> $\mt{not\ defeated}(s,\non A)$.
\end{clause}
\begin{clause}%\label{defeated}
  $\mt{defeated}(s,\non A)$\ :-\\
  \> $\mt{body}_t^d(args(A))$.
\end{clause}

For example, clauses derived from $t$ in the previous example are now
\[
\begin{array}{lrcl}
 &  \mt{lambda}(q(X, Y))  & \mbox{:-} &  \\
 &&& \mt{not\ definitely}(not\_q(X, Y)), \\
 &&& \mt{body}_t^\lambda(X, Y). \\
 
& \mt{defeasibly}(q(X, Y))  & \mbox{:-} &  \\
 &&& \mt{not\ definitely}(not\_q(X, Y)),  \\
 &&& \mt{body}_t^d(X, Y), \\
 &&&  \mt{not\ overruled}(q(X, Y)). \\
 
 &  \mt{overruled}(not\_q(X, Y))  & \mbox{:-} &  \\
 &&& \mt{body}_t^\lambda(X, Y), \\
 &&& \mt{not\ defeated}(t, q(X, Y)). \\
 
 & \mt{defeated}(s, not\_q(X, Y)) & \mbox{:-} &  \\
 &&& \mt{body}_t^d(X, Y)). \\
\end{array}
\]

Next,
we unfold clauses \ref{lambda1} and \ref{defeasibly1}  using the clauses for $\mt{definitely}$.
This results in additional clauses (including unit clauses) for $\mt{lambda}$, and $\mt{defeasibly}$
corresponding to the existing clauses for $\mt{definitely}$.

At this stage,
the transformed program consists of the following clauses.
Recall that we use $\ARROW$ to range over the different kinds of arrows used in a defeasible theory.

\noindent
The introduced clauses \ref{body:lam}--\ref{body:strict} remain.

\noindent
For every fact $F$ in $D$, we have the unit clauses $\mt{definitely}(F)$, $\mt{lambda}(F)$, and $\mt{defeasibly}(F)$ in the program.

\noindent
For every strict rule
$
\begin{array}{lrcl}
r: &   B_1, \ldots, B_n       & \rightarrow & A \\
\end{array}
$
in $D$, we have the clauses
\begin{Clause}  \label{definitely3}
  $\mt{definitely}(A)$\ :-\\
  \> $\mt{body}_r^\Delta(args(A))$.
\end{Clause}
\begin{Clause}   \label{c:lambda1}
  $\mt{lambda}(A)$\ :-\\
  \> $\mt{body}_r^\Delta(args(A))$.
\end{Clause}
\begin{Clause} 
  $\mt{defeasibly}(A)$\ :-\\
  \> $\mt{body}_r^\Delta(args(A))$.
\end{Clause}
For every strict or defeasible rule
$
\begin{array}{lrcl}
r: &   B_1, \ldots, B_n       & \ARROW & A \\
\end{array}
$
in $D$, we have the clause
\begin{Clause}  \label{c:lambda2}
  $\mt{lambda}(A)$\ :-\\
  \> $\mt{not\ definitely}(\non A)$,\\
  \>  $\mt{body}_r^\lambda(args(A))$.
\end{Clause}
and the clause
\begin{Clause}%\label{defeasibly2}
  $\mt{defeasibly}(A)$\ :-\\
  \> $\mt{not\ definitely}(\non A)$,\\
  \> $\mt{body}_r^d(args(A))$,\\
  \> $\mt{not\ overruled}(A)$.
\end{Clause}
For every rule
$
\begin{array}{lrcl}
s: &   B_1, \ldots, B_n       & \ARROW & \non A \\
\end{array}
$
in $D$, we have the clause
\begin{Clause}%\label{overruled}
  $\mt{overruled}(A)$\ :-\\
  \>  $\mt{body}_s^\lambda(args(A))$,\\
  \> $\mt{not\ defeated}(s,\non A)$.
\end{Clause}
For every strict or defeasible rule
$
\begin{array}{lrcl}
t: &   B_1, \ldots, B_n       & \ARROW & A \\
\end{array}
$
in $D$ that is superior to a rule $s$ for $\non A$, we have the clause
\begin{Clause}   \label{defeated3}
  $\mt{defeated}(s,\non A)$\ :-\\
  \> $\mt{body}_t^d(args(A))$.
\end{Clause}

We now eliminate the function symbols in the clauses.
For each predicate $p$ in $D$, of arity $n$, we introduce the clauses
\begin{Clause}  \label{conv1}
  $\mt{definitely\_p}(X_1,\ldots,X_n)$\ :-\\
  \> $\mt{definitely}(p(X_1,\ldots,X_n))$.
\end{Clause}
\begin{Clause}
  $\mt{definitely\_not\_p}(X_1,\ldots,X_n)$\ :-\\
  \> $\mt{definitely}(not\_p(X_1,\ldots,X_n))$.
\end{Clause}
\begin{Clause}
  $\mt{lambda\_p}(X_1,\ldots,X_n)$\ :-\\
  \> $\mt{lambda}(p(X_1,\ldots,X_n))$.
\end{Clause}
\begin{Clause}
  $\mt{lambda\_not\_p}(X_1,\ldots,X_n)$\ :-\\
  \> $\mt{lambda}(not\_p(X_1,\ldots,X_n))$.
\end{Clause}
\begin{Clause}
  $\mt{defeasibly\_p}(X_1,\ldots,X_n)$\ :-\\
  \> $\mt{defeasibly}(p(X_1,\ldots,X_n))$.
\end{Clause}
\begin{Clause}
  $\mt{defeasibly\_not\_p}(X_1,\ldots,X_n)$\ :-\\
  \> $\mt{defeasibly}(not\_p(X_1,\ldots,X_n))$.
\end{Clause}
\begin{Clause}
  $\mt{overruled\_p}(X_1,\ldots,X_n)$\ :-\\
  \> $\mt{overruled}(p(X_1,\ldots,X_n))$.
\end{Clause}
\begin{Clause}
  $\mt{overruled\_not\_p}(X_1,\ldots,X_n)$\ :-\\
  \> $\mt{overruled}(not\_p(X_1,\ldots,X_n))$.
\end{Clause}
\begin{Clause}
  $\mt{defeated\_p}(S, X_1,\ldots,X_n)$\ :-\\
  \> $\mt{defeated}(S, p(X_1,\ldots,X_n))$.
\end{Clause}
\begin{Clause}   \label{conv10}
  $\mt{defeated\_not\_p}(S, X_1,\ldots,X_n)$\ :-\\
  \> $\mt{defeated}(S, not\_p(X_1,\ldots,X_n))$.
\end{Clause}

We now fold every body literal (except those in the above clauses)
in the program containing a function symbol by the appropriate clause (\ref{conv1}--\ref{conv10}).
This is essentially the replacement of atoms involving a function symbol
with an atom with a predicate name incorporating the function symbol.
Then we unfold the body atom of each of the above clauses (\ref{conv1}--\ref{conv10}).
There are now no occurrences of the original predicates in $\M(D)$
($\mt{definitely}$, $\mt{lambda}$, $\mt{defeasibly}$, $\mt{overruled}$ and $\mt{defeated}$)
in the bodies of rules.
We denote the resulting program by $T_D$.

The introduced predicates, such as $\mt{defeasibly\_p}$, together fully represent the original predicates, such as 
$\mt{defeasibly}$ in $T_D$.

\begin{propn}   \label{prop:conv}
For every literal $q(\vec{a})$,
\begin{itemize}
\item
$T_D \models_{WF}  \mt{definitely}(q(\vec{a}))$ iff $T_D \models_{WF}  \mt{definitely\_q}(\vec{a})$
\item
$T_D \models_{WF}  \mt{lambda}(q(\vec{a}))$ iff $T_D \models_{WF}  \mt{lambda\_q}(\vec{a})$
\item
$T_D \models_{WF}  \mt{defeasibly}(q(\vec{a}))$ iff $T_D \models_{WF}  \mt{defeasibly\_q}(\vec{a})$
\end{itemize}
\end{propn}
\skipit{
\begin{proof}
These statements are true once the clauses \ref{conv1}--\ref{conv10} are introduced, by inspection of those clauses.
The subsequent transformations preserve the semantics of the program.
\end{proof}
}

\subsection{The Resulting Program}

In the following, $\mt{q}$ (possibly subscripted) may have the form $\mt{p}$ or $\mt{not\_p}$.
We write $\mt{\non q}$ as part of a predicate name to represent, respectively, $\mt{not\_p}$ or $\mt{p}$.
We can now outline what the final program $T_D$ looks like.

For every fact $q(\vec{a})$ in $D$, there are unit clauses $\mt{definitely\_q}(\vec{a})$, $\mt{lambda\_q}(\vec{a})$, and \linebreak $\mt{defeasibly\_q}(\vec{a})$.

For every strict rule
$
\begin{array}{lrcl}
r: &   q_1(\vec{a_1}), \ldots, q_n(\vec{a_n})       & \rightarrow & q(\vec{a}) \\
\end{array}
$
in $D$, we have the clauses
\begin{Clause}   \label{final:definitely}
  $\mt{definitely\_q}(\vec{a})$\ :-\\
  \> $\mt{body}_r^\Delta(\vec{a})$.
\end{Clause}
\begin{Clause}
  $\mt{lambda\_q}(\vec{a})$\ :-\\
  \> $\mt{body}_r^\Delta(\vec{a})$.
\end{Clause}
\begin{Clause}
  $\mt{defeasibly\_q}(\vec{a})$\ :-\\
  \> $\mt{body}_r^\Delta(\vec{a})$.
\end{Clause}
\begin{Clause}
  $\mt{body}_r^\Delta(\vec{a})$\ :-\\
  \> $\mt{definitely\_q_1}(\vec{a_1})$, \ldots, $\mt{definitely\_q_n}(\vec{a_n})$.
\end{Clause}
For every strict or defeasible rule
$
\begin{array}{lrcl}
r: &   q_1(\vec{a_1}), \ldots, q_n(\vec{a_n})       & \ARROW & q(\vec{a}) \\
\end{array}
$
in $D$, we have the clauses
\begin{Clause}  \label{final:lambda2}
  $\mt{lambda\_q}(\vec{a})$\ :-\\
  \> $\mt{not\ definitely\_\non q}(\vec{a})$,\\
  \>  $\mt{body}_r^\lambda(\vec{a})$.
\end{Clause}
\begin{Clause}  \label{final:defeasibly2}
  $\mt{defeasibly\_q}(\vec{a})$\ :-\\
  \> $\mt{not\ definitely\_\non q}(\vec{a})$,\\
  \> $\mt{body}_r^d(\vec{a})$,\\
  \> $\mt{not\ overruled\_q}(\vec{a})$.
\end{Clause}
\begin{Clause}   \label{final:lambda_body}
  $\mt{body}_r^\lambda(\vec{a})$\ :-\\
  \> $\mt{lambda\_q_1}(\vec{a}_1)$, \dots, $\mt{lambda\_q_n}(\vec{a}_n)$.
\end{Clause}

For every rule
$
\begin{array}{lrcl}
s: &   q_1(\vec{a_1}), \ldots, q_n(\vec{a_n})       & \ARROW &  q(\vec{a}) \\
\end{array}
$
in $D$, we have the clauses
\begin{Clause}   \label{body_def}
  $\mt{body}_r^d(\vec{a})$\ :-\\
  \> $\mt{defeasibly\_q_1}(\vec{a}_1)$, \dots, $\mt{defeasibly\_q_n}(\vec{a}_n)$.
\end{Clause}
\begin{Clause}  \label{final:overruled}
  $\mt{overruled\_\non q}(\vec{a})$\ :-\\
  \>  $\mt{body}_s^\lambda\_q(\vec{a})$,\\
  \> $\mt{not\ defeated\_q}(s, \vec{a})$.
\end{Clause}

For every strict or defeasible rule
$
\begin{array}{lrcl}
t: &   q_1(\vec{a_1}), \ldots, q_n(\vec{a_n})       & \ARROW & q(\vec{a}) \\
\end{array}
$
in $D$ that is superior to a rule $s$ for $\non q$, we have the clause
\begin{Clause}   \label{final:defeated}
  $\mt{defeated\_\non q}(s, \vec{a})$\ :-\\
  \> $\mt{body}_t^d(\vec{a})$.
\end{Clause}

The transformed program $T_D$  also contains clauses for
$\mt{definitely}$, $\mt{lambda}$, $\mt{defeasibly}$, \linebreak
$\mt{overruled}$ and $\mt{defeated}$
(the original predicates)
which are not needed for computation of the consequences of $D$ in $\DL(\pl)$
but are convenient to prove correctness of the result.
These consist of unit clauses 
$\mt{definitely}(F)$, $\mt{lambda}(F)$, and $\mt{defeasibly}(F)$, for each fact $F$ in $D$,
and clauses \ref{definitely3}--\ref{defeated3}.
These clauses will be deleted shortly.

The size of $T_D$ is almost clearly linear in the size of $D$.\footnote{
We measure size by the number of symbols in a defeasible theory or logic program.
}
Almost every clause is derived from a rule of $D$ (or fact) and is linear in the size of that rule.
Further, each rule gives rise to at most 9 clauses
(excluding clauses of the form \ref{final:defeated}). 
The only problems are the clauses of the form \ref{defeated3} and \ref{final:defeated}.
There is one of each such clause for each superiority statement $s < t$,
but the size of these clauses is not necessarily constant;
the size of the arguments $\vec{a}$ is bounded by the size of $s$ and $t$, but that is not constant.
We need to address such clauses more carefully.

For each predicate $p$ in $D$,
let $D_p$ be the restriction of $D$ to rules defining $p$ and $\neg p$.
Let $K_p$ be the number of distinct superiority statements in $D$ about rules for $p$ and $\neg p$,
$M_p$ be the number of rules involved in those superiority statements,
and $A_p$ be the arity of $p$.
Then $K_p+1 \leq M_p \leq 2K_p$, so O($K_p$) = O($M_p$).
$D_p$ must contain at least $M_p$ rules, each of size greater than $A_p$.
That is, $M_pA_p < size(D_p)$.
The size of the clauses in $T_D$ derived from \ref{final:defeated} 
is bounded by the product of $K_p$ (the number of clauses) and a linear term in $A_p$ (the size of each clause).
It follows that the size of the clauses derived from \ref{final:defeated} is linear in the size of $D_p$.
Thus, the total size of all clauses in $T_D$ derived from \ref{final:defeated} is 
O($\sum_{p \in \Pi}  K_pA_p$) $\leq$ O($\sum_{p \in \Pi}  size(D_p)$) $\leq$ O($size(D)$).

Thus the modifications can lead only to a linear blow-up from $D$ to $T_D$.
\begin{propn}  \label{prop:linear}
The size of the resulting program $T_D$ is linear in the size of $D$.
\end{propn}

Let us consider now the earlier example of rules in $D$.
\begin{example}
Consider a defeasible theory consisting of the rules
\[
\begin{array}{lrcl}
s: &   p(X, Y), q(Y, X)               & \Rightarrow & \neg q(X, Y) \\
t: &   p(X, Z), \neg p(Z, Y)        & \Rightarrow & \phantom{\neg} q(X, Y) \\
\end{array}
\]
with $t > s$, and some facts for $p$.

The transformed program contains
\[
\begin{array}{lrcl}
 &  \mt{lambda\_q}(X, Y)  & \mbox{:-} &  \\
 &&& \mt{not\ definitely\_not\_q}(X, Y), \\
 &&& \mt{body}_t^\lambda(X, Y). \\
 
& \mt{defeasibly\_q}(X, Y)  & \mbox{:-} &  \\
 &&& \mt{not\ definitely\_not\_q}(X, Y),  \\
 &&& \mt{body}_t^d(X, Y), \\
 &&&  \mt{not\ overruled\_q}(X, Y). \\
 
 &  \mt{overruled\_not\_q}(X, Y)  & \mbox{:-} &  \\
 &&& \mt{body}_t^\lambda(X, Y), \\
 &&& \mt{not\ defeated\_q}(t, X, Y). \\
 
 & \mt{defeated\_not\_q}(s, X, Y) & \mbox{:-} &  \\
 &&& \mt{body}_t^d(X, Y). \\

 &  \mt{body}_t^\lambda(X,Y) & \mbox{:-} &  \\
 &&& \mt{lambda\_p}(X, Z), \mt{lambda\_not\_q}(Z, Y). \\
  
 &  \mt{body}_t^d(X,Y) & \mbox{:-} &  \\
 &&& \mt{defeasibly\_p}(X, Z), \mt{defeasibly\_not\_q}(Z, Y). \\

 &  \mt{lambda\_not\ _q}(X, Y)  & \mbox{:-} &  \\
 &&& \mt{not\ definitely\_q}(X, Y), \\
 &&& \mt{body}_s^\lambda(X, Y). \\
 
& \mt{defeasibly\_not\_q}(X, Y)  & \mbox{:-} &  \\
 &&& \mt{not\ definitely\_q}(X, Y),  \\
 &&& \mt{body}_s^d(X, Y), \\
 &&&  \mt{not\ overruled\_q}(X, Y). \\
 
 &  \mt{overruled\_q}(X, Y)  & \mbox{:-} &  \\
 &&& \mt{body}_s^\lambda(X, Y), \\
 &&& \mt{not\ defeated\_not\_q}(s, X, Y). \\ 
 
 &  \mt{body}_s^\lambda(X,Y) & \mbox{:-} &  \\
 &&& \mt{lambda\_p}(X, Y), \mt{lambda\_q}(Y, X). \\
  
 &  \mt{body}_s^d(X,Y) & \mbox{:-} &  \\
 &&& \mt{defeasibly\_p}(X, Y), \mt{defeasibly\_q}(Y, X). \\
 
\end{array}
\]
The first six clauses are derived from $t$, and the last five from $s$.
\end{example}

Notice that $\DL(\pl)$ as defined in Section~\ref{sec:sdl} only refers to positive consequences.
As a result, it is only the positive consequences of $\M_{\pl}(D)$ for the predicates
$\mt{defeasibly}$, $\mt{definitely}$ and, to a lesser extent, $\mt{lambda}$
that we must be concerned with.
This extends to the transformed program.

$\M_{\pl}(D)$ and the transformed program $T_D$ are equivalent under the well-founded semantics.

\begin{propn}    \label{prop:trans}
Let $D$ be a defeasible theory, and
let $T_D$ be the transformed version of $\M_{\pl}(D)$.
Let $q$ be a literal.
\begin{itemize}
\item
$\M_{\pl}(D) \models_{WF}  \mt{definitely}(q)$ iff $T_D \models_{WF}  \mt{definitely}(q)$
\item
$\M_{\pl}(D) \models_{WF}  \mt{lambda}(q)$ ~~~~~~~\, iff $T_D \models_{WF}  \mt{lambda}(q)$
\item
$\M_{\pl}(D) \models_{WF}  \mt{defeasibly}(q)$ iff $T_D \models_{WF}  \mt{defeasibly}(q)$
\end{itemize}
\end{propn}
\skipit{
\begin{proof}
$\M(D)$ is transformed to $T_D$ by a series of unfolding and folding transformations
and additions of clauses defining new predicates,
that preserve the well-founded semantics \cite{unstable,Aravindan}.
\end{proof}
}

Similar results apply for $\M_{\pl^*}(D)$.  
The structure of the two metaprograms is largely the same,
with minor variations in clauses \ref{defeasibly2_indiv} and \ref{overruled_indiv},
and a substantial simplification in \ref{defeated_indiv}.
Thus the same transformations apply, except to clauses derived from  \ref{defeated_indiv}.
The resulting program contains clauses as described in
\ref{final:definitely}--\ref{final:lambda2}, \ref{final:lambda_body}--\ref{body_def},
as well as clauses defining $\mt{defeasibly\_q}$, $\mt{overruled\_\non q}$, and $\mt{defeated\_q}$.

Reflecting the minor difference between \ref{defeasibly2_indiv} and \ref{defeasibly2},
the transformed $\M_{\pl^*}(D)$
contains clauses like \ref{final:defeasibly2_indiv} rather than \ref{final:defeasibly2}.
Similarly, \ref{final:overruled_indiv} is only a minor variation of \ref{final:overruled},
while 
\ref{final:defeated_indiv}
is a set of unit clauses $\mt{defeated}(r, s)$ 
corresponding to the superiority statements $r > s$ in $D$.

\begin{Clause}   \label{final:defeasibly2_indiv}
  $\mt{defeasibly\_q}(\vec{a})$\ :-\\
  \> $\mt{not\ definitely\_\non q}(\vec{a})$,\\
  \> $\mt{body}_r^d(\vec{a})$,\\
  \> $\mt{not\ overruled\_q}(r, \vec{a})$.
\end{Clause}
\begin{Clause}  \label{final:overruled_indiv}
  $\mt{overruled\_\non q}(r, \vec{a})$\ :-\\
  \>  $\mt{body}_s^\lambda\_q(\vec{a})$,\\
  \> $\mt{not\ defeats\_q}(r, s)$.
\end{Clause}
\begin{Clause}  \label{final:defeated_indiv}
  $\mt{defeats\_q}(r, s)$.
\end{Clause}

The size of the transformed program is clearly linear in the size of $D$
since the only problematic clauses \ref{defeated3} and \ref{final:defeated} in $T_D$ for $\M_{\pl}(D)$ 
are unit clauses in the transformed program for $\M_{\pl^*}(D)$.

With these variations, we have similar results to
Propositions \ref{prop:trans} and \ref{prop:linear}.

\begin{propn}    \label{prop:trans*}
Let $D$ be a defeasible theory, and
let $T^*_D$ be the transformed version of $\M_{\pl^*}(D)$.
Let $q$ be a literal.
\begin{itemize}
\item
$\M_{\pl^*}(D) \models_{WF}  \mt{definitely}(q)$ iff $T^*_D \models_{WF}  \mt{definitely}(q)$
\item
$\M_{\pl^*}(D) \models_{WF}  \mt{lambda}(q)$  ~~~~~~~\:  iff $T^*_D \models_{WF}  \mt{lambda}(q)$
\item
$\M_{\pl^*}(D) \models_{WF}  \mt{defeasibly}(q)$ \! iff $T^*_D \models_{WF}  \mt{defeasibly}(q)$
\item
The size of $T^*_D$ is linear in the size of $D$.
\end{itemize}
\end{propn}

Combining the previous results, we summarise the relationship between
the applied metaprogram and the compiled version as follows.
Let $S_D$ be the transformed program $T_D$ after deleting the clauses for 
$\mt{definitely}$, $\mt{lambda}$, $\mt{defeasibly}$, $\mt{overruled}$ and $\mt{defeated}$,
and let $S^*_D$ be $T^*_D$ after deleting similar clauses.

\begin{theorem}    \label{thm:trans_all}
Let $D$ be a defeasible theory and $S_D$ ($S^*_D$) be the transformed program.
Let $q(\vec{a})$ be a ground literal.
$q$ has the form $p$ or $\neg p$ in $D$, but
$\mt{p}$ or $\mt{not\_p}$ in the transformed metaprogram.
\begin{itemize}
\item
$\M_{\pl}(D) \models_{WF}  \mt{definitely}(q(\vec{a}))$ iff  $S_D \models_{WF} \mt{definitely\_q}(\vec{a})$
\item
$\M_{\pl}(D) \models_{WF}  \mt{lambda}(q(\vec{a}))$  ~~~~~~~\: iff  $S_D \models_{WF}  \mt{lambda\_q}(\vec{a})$
\item
$\M_{\pl}(D) \models_{WF}  \mt{defeasibly}(q(\vec{a}))$ iff  $S_D \models_{WF}  \mt{defeasibly\_q}(\vec{a})$
\smallskip
\item
$\M_{\pl^*}(D) \models_{WF}  \mt{definitely}(q(\vec{a}))$ iff  $S^*_D \models_{WF} \mt{definitely\_q}(\vec{a})$
\item
$\M_{\pl^*}(D) \models_{WF}  \mt{lambda}(q(\vec{a}))$  ~~~~~~~\: iff  $S^*_D \models_{WF}  \mt{lambda\_q}(\vec{a})$
\item
$\M_{\pl^*}(D) \models_{WF}  \mt{defeasibly}(q(\vec{a}))$ iff  $S^*_D \models_{WF}  \mt{defeasibly\_q}(\vec{a})$
\end{itemize}
\end{theorem}
\skipit{
\begin{proof}
By Propositions \ref{prop:trans1} and \ref{prop:trans},
$\M_{\pl}(D)$ is equivalent to $P_D$ and $T_D$
on the predicates  $\mt{definitely}$, $\mt{lambda}$, and $\mt{defeasibly}$.
Similarly, $\M_{\pl^*}(D)$ is equivalent to $P^*_D$ and $T^*_D$, by Proposition \ref{prop:trans*}.
By Proposition \ref{prop:conv}, literals such as $\mt{defeasibly\_q}(\vec{a})$ are inferred iff $\mt{defeasibly}(q(\vec{a}))$ is inferred in $T_D$ (and similarly for $T^*_D$).
Finally, the deletion of predicates from $T_D$ to get $S_D$ (and from $T^*_D$ to get $S^*_D$)
preserves the equivalence because the predicates of interest in $S_D$ do not depend on the deleted predicates.
\end{proof}
}

Additionally, following Proposition \ref{prop:linear},
$S_D$ and $S^*_D$ are linear in the size of $D$.

The results in this section depend only on the correctness of the transformations used.
Consequently, they extend beyond the well-founded semantics to many other logic programming semantics.
Indeed, 
\cite{Aravindan} showed that these transformations preserve the
regular models \cite{regular}, stable theory semantics \cite{stable_theory},
and
stable semantics \cite{stable}, as well as the well-founded semantics.
\cite{cdr} extended this approach to
partial stable models \cite{Pmodels} and L-stable models \cite{Lstable}.
Earlier work showed that these transformations preserved
the (two-valued) Clark-completion semantics \cite{trans} 
and Fitting's and Kunen's semantics \cite{BossiCE92}.
However, these latter semantics do not express the ordering of computation
(that is, $P_\Delta$ then $P_\lambda$ then $P_{\pl}$).

The results are also independent of whether $D$ is function-free or not.
Moreover, if constraints (in the sense of constraint logic programming \cite{JM94})
are permitted in the defeasible theory, they can be expressed in the corresponding CLP language,
and the results still apply.

Finally, a similar sequence of transformations would apply to the metaprogram of almost any (sensible)
defeasible logic.
Thus this approach provides a provably correct compilation of defeasible logics to \Datalog{} 
using only the metaprogram representation and a simple fold/unfold transformation system.

This section has established the correctness of the mapping from the metaprogram to \Datalog{}.
We now turn to establishing the correctness of the metaprogram.

% { $\M(D) $ partial eval  $P_D$  transform  $T_D$  delete old preds  $S_D$  }

\section{Correctness of the Metaprogram}  \label{sec:semantics}

To establish the correctness of the compilation of $\DL(\pl)$ (and $\DL(\pl^*)$) to \Datalog{},
we need to verify that the metaprogram presented in Section \ref{sec:metaprogram}
is correct with respect to the proof theory defined in Section \ref{sec:sdl}.

In general, $D$ is not propositional, but the inference rules in Section \ref{sec:sdl}
are formulated for propositional defeasible theories, so we consider $D$ to be a schema defining
the sets of ground instances of rules in $D$.
As a result, the atoms are essentially propositional and the inference rules can be applied.
Under the well-founded semantics, a logic program and the ground instances of all its clauses are equivalent.
As a result, in the following proofs we consider only ground rules and ground clauses.

The proofs for $\Delta$ and $\lambda$ are straightforward inductions.

\begin{theorem}    \label{thm:correct_deflam}
Let $D$ be a defeasible theory and $q$ be a ground literal.
$q$ has the form $p(\vec{a})$ or $\neg p(\vec{a})$ in $D$, but
$\mt{p}(\vec{a})$ or $\mt{not\_p}(\vec{a})$ in $\M_{\pl}(D)$.
\begin{itemize}
\item
$q \in P_\Delta$  
   iff  $\M_{\pl}(D) \models_{WF} \mt{definitely}(q)$
\item
$q \in P_\lambda$
\! iff  $\M_{\pl}(D) \models_{WF}  \mt{lambda}(q)$
\end{itemize}
\end{theorem}
\skipit{
\begin{proof}
It is convenient to prove this from the program $P_D$, rather than $\M_{\pl}(D)$.
As shown in Proposition \ref{prop:trans}, these two programs are equivalent for the predicates of interest.
For simplicity of notation, we write $P$ instead of $ground(P_D)$.

Part 1 $\Rightarrow$ \\
The proof is by induction on the length of a proof for $q$, with induction hypothesis:
if $+\Delta q$ has proof of length $\leq n$ then $\mt{definitely(q)} \in \lfp(\W_P)$.
For $n=1$, if $+\Delta q$ has a proof of length 1 then either $q$ is a fact or 
$q$ is the head of a strict rule with an empty body.
In either case,  $\mt{definitely(q)}$ is a unit clause in $P$, so $\mt{definitely(q)} \in \lfp(\W_P)$.
If  $+\Delta q$ has a proof of length $n+1$,
then there is an instance $q_1,\ldots,q_k \rightarrow q$ of a strict rule in $D$
such that each $q_i$ has a proof of length $\leq n$.
By the induction hypothesis, $\mt{definitely(q_i)} \in \lfp(\W_P)$, for each $i$.
$P$ must contain a clause 
$$
\mt{definitely(q)}  \mbox{~~:-~~} \mt{definitely(q_1)}, \ldots, \mt{definitely(q_k)}
$$
from the definition of $P$.
Hence $\mt{definitely(q)} \in \lfp(\W_P)$.
Thus, by induction, if $+\Delta q$ has a proof then $\mt{definitely(q)} \in \lfp(\W_P)$.

Part 1 $\Leftarrow$ \\
The proof is by induction on the length of the Kleene sequence, with induction hypothesis:
if $\mt{definitely(q)} \in \W_P \uparrow n$ then $q \in P_\Delta$.
If $\mt{definitely(q)} \in \W_P \uparrow 1$ then $\mt{definitely(q)}$ is a unit clause in $P$.
It follows that $q$ is either a fact or the head of a strict rule with empty body in $D$.
Consequently,  $q \in P_\Delta$.
If $\mt{definitely(q)} \in \W_P \uparrow( n+1)$ then $P$ must have a clause with head $\mt{definitely(q)}$, say
$$
\mt{definitely(q)}  \mbox{~~:-~~} \mt{definitely(q_1)}, \ldots, \mt{definitely(q_k)}
$$
where $\mt{definitely(q_i)} \in \W_P \uparrow n$, for $i = 1,\ldots,k$.
By  the induction hypothesis, $+\Delta q_i \in P_\Delta$, for $i = 1,\ldots,k$.
Hence $+\Delta q \in P_\Delta$.

Part 2 $\Rightarrow$ \\
The proof is by induction on the length of a proof for $q$, with induction hypothesis:
if $+\lambda q$ has proof of length $\leq n$ then $\mt{lambda(q)} \in \lfp(\W_P)$.
For $n=1$, if $+\lambda q$ has a proof of length 1 then either $q$ is a fact or 
$q$ is the head of a strict or defeasible rule with an empty body.
In either case,  $\mt{lambda(q)}$ is a unit clause in $P$, so $\mt{lambda(q)} \in \lfp(\W_P)$.
If  $+\lambda q$ has a proof of length $n+1$,
then there is an instance $q_1,\ldots,q_k \rightarrow q$ of a strict or defeasible rule in $D$
such that each $q_i$ has a proof of length $\leq n$ and $+\Delta \non q \notin P_\Delta$.
By the induction hypothesis, $\mt{lambda(q_i)} \in \lfp(\W_P)$, for each $i$.
By the first part of this theorem, $\mt{definitely(\non q)} \notin \lfp(\W_P)$,
and hence $not~\mt{definitely(\non q)} \in \lfp(\W_P)$
(because the well founded semantics is total on stratified downward-closed subprograms
like the clauses defining $\mt{definitely}$).
$P$ must contain a clause 
$$
\mt{lambda(q)}  \mbox{~~:-~~} not ~ \mt{definitely(\non q)}, \mt{lambda(q_1)}, \ldots, \mt{lambda(q_k)}
$$
from the definition of $P$.
Hence $\mt{lambda(q)} \in \lfp(\W_P)$.
Thus, by induction, if $+\lambda q$ has a proof then $\mt{lambda(q)} \in \lfp(\W_P)$.

Part 2 $\Leftarrow$ \\
The proof is by induction on the length of the Kleene sequence, with induction hypothesis:
if $\mt{lambda(q)} \in \W_P \uparrow n$ then $q \in P_\lambda$.
If $\mt{lambda(q)} \in \W_P \uparrow 1$ then $\mt{lambda(q)}$ is a unit clause in $P$.
It follows that $q$ is either a fact or the head of a strict or defeasible rule with empty body in $D$.
Consequently,  $q \in P_\lambda$.
If $\mt{lambda(q)} \in \W_P \uparrow( n+1)$ with $n>0$, then $P$ must have a clause with head $\mt{lambda(q)}$, say
$$
\mt{lambda(q)}  \mbox{~~:-~~}  not ~ \mt{definitely(\non q)}, \mt{lambda(q_1)}, \ldots, \mt{lambda(q_k)}
$$
where $\mt{lambda(q_i)} \in \W_P \uparrow n$, for $i = 1,\ldots,k$
and $not ~ \mt{definitely(\non q)} \in \lfp(\W_P)$.
By  the induction hypothesis, $+\lambda q_i \in P_\lambda$, for $i = 1,\ldots,k$
and, by the first part of this theorem, $+\Delta q \notin P_\Delta$.
$D$ must have a rule $q_1,\ldots,q_k \ARROW q$, where $\ARROW$ is $\rightarrow$ or $\Rightarrow$,
from which the clause in $P$ mentioned above arises.
Thus part (2) of the inference rule for $\lambda$ applies, and $+\lambda q$ can be proved.
Hence $+\lambda q \in P_\lambda$.
Thus, by induction, if $\mt{lambda(q)} \in \lfp(\W_P)$ then $+\lambda q \in P_\lambda$.

\end{proof}
}

As a consequence of this theorem and Theorem \ref{thm:trans_all},
the compilation is correct with respect to the inference rules $+\Delta$ and $+\lambda$.

\begin{theorem}    \label{thm:correct2?}
Let $D$ be a defeasible theory and $S_D$ ($S^*_D$) be the transformed program.
Let $q(\vec{a})$ be a ground literal.
$q$ has the form $p$ or $\neg p$ in $D$, but
$\mt{p}$ or $\mt{not\_p}$ in the transformed metaprogram.
\begin{itemize}
\item
$q \in P_\Delta$ iff  $S_D \models_{WF} \mt{definitely\_q}(\vec{a})$
iff $S^*_D \models_{WF} \mt{definitely\_q}(\vec{a})$
\item
$q \in P_\lambda$ \ iff  $S_D \models_{WF}  \mt{lambda\_q}(\vec{a})$ \hspace{0.65cm}
 iff  $S^*_D \models_{WF}  \mt{lambda\_q}(\vec{a})$
\end{itemize}
\end{theorem}

Demonstrating the correctness of the compilation for the main tags is more difficult,
largely because of the greater complexity of the inference rules.
We will use the following lemma to structure the proof.

\begin{lemma}  \label{lemma:corr}
Let $(L_1, \leq_1)$ and $(L_2, \leq_2)$ be partial orders.
Let $\Psi : L_1 \rightarrow L_2$ and $\Gamma : L_2 \rightarrow L_1$ be monotonic functions.
Let $X_1 \in L_1$ and $X_2 \in L_2$.

If the following conditions hold
\begin{enumerate}
\item  \label{pt:psi}
$X_2 \leq_2 \Psi(X_1)$ 
\item  \label{pt:gamma}
$X_1 \leq_1 \Gamma(X_2)$
\item  \label{pt:eq}
$\Psi(\Gamma(X_2)) =X_2$
\end{enumerate}
then $X_2 = \Psi(X_1)$. 
\end{lemma}
\skipit{
\begin{proof}
$\Psi(X_1) \leq_2 \Psi(\Gamma(X_2)) = X_2 \leq_2 \Psi(X_1)$,
using monotonicity of $\Psi$ and the conditions of the theorem.
Hence $X_2 = \Psi(X_1)$.
\end{proof}
}

In particular,
let $f_1 : L_1 \rightarrow L_1$ and $f_2 : L_2 \rightarrow L_2$ be monotonic functions,
and let $(L_1, \leq_1)$ and $(L_2, \leq_2)$ be complete partial orders,
so that
$X_1 = \lfp(f_1)$ and $X_2 = \lfp(f_2)$ exist.
Then, under the conditions of this lemma, $\Psi(\lfp(f_1)) = \lfp(f_2)$

In the application of the lemma, 
$(L_1, \leq_1)$ is the set of Herbrand interpretations of $P_D$ under the containment ordering and
$(L_2, \leq_2)$ is the set of sets of tagged literals from $D$, again  under the containment ordering.
$f_1$ is $\W_{P_D}$, so
$X_1$ is the well-founded model of $P_D$ and
$f_2$ is the function that applies the inference rules of $\DL(\pl)$ in every way possible,
so $X_2$ is $P_{\pl} \cup P_{\lambda} \cup P_{\Delta}$,
the least deductively-closed set under the $\DL(\pl)$ inference rules.

\begin{theorem}    \label{thm:correct_defs}
Let $D$ be a defeasible theory and $q$ be a ground literal.
$q$ has the form $p(\vec{a})$ or $\neg p(\vec{a})$ in $D$, but
$\mt{p}(\vec{a})$ or $\mt{not\_p}(\vec{a})$ in $\M_{\pl}(D)$.
\begin{itemize}
\item
$D \vdash +\pl q$ iff  $\M_{\pl}(D) \models_{WF}  \mt{defeasibly}(q)$
\item
$D \vdash +\pl^* q$ iff  $\M_{\pl^*}(D) \models_{WF}  \mt{defeasibly}(q)$
\end{itemize}
\end{theorem}
\skipit{
\begin{proof}
It is convenient to prove this from the program $P_D$, rather than $\M_{\pl}(D)$.
As shown in Proposition \ref{prop:trans}, these two programs are equivalent for the predicates of interest.
For simplicity of notation, we write $P$ instead of $ground(P_D)$.

Let $W$ denote the well founded model of $P$ as derived in Section \ref{sec:trans}.
Let $X = \Psi(W)$, where $\Psi(W)$ is defined as $\{ +\Delta q~|~ \mt{definitely(q)} \in W \} \cup
\{ +\lambda q~|~ \mt{lambda(q)} \in W \}  \cup
\{ +\pl q~|~ \mt{defeasibly(q)} \in W \}$.
Clearly $\Psi$ is monotonic.
Note that, by Proposition \ref{prop:trans1}, $W$ is also the well founded model of $\M_{\pl}(D)$ as derived in Section \ref{sec:trans}.

We claim that $X$ is $\pl$-deductively closed  from $D$.
Consider a literal $p$, and suppose that $+\pl p$ can be inferred from $X$.
Then either (1) $\PD{p} \in X$ or
(2) there is a strict or defeasible rule $r$ where all body literals $p_i$ are tagged by $\pl$ in $X$;
$\PD{\non p} \notin X$; and
for every rule $s$ for $\non p$ either
some body literal $q_i$ has $+\lambda q_i \notin X$, or
there is a strict or defeasible rule $t$ where $t > s$ and all body literals $p'_i$ are tagged by $\pl$ in $X$.
If (1) then we must have $ \mt{definitely(p)} \in W$.
But then, since $W$ is a model of $P$, $\mt{defeasibly(p)} \in W$ and hence $+\pl p$ is in $X$ (by the definition of $X$).

If (2) then 
(2.1) $\mt{defeasibly(p_i)} \in W$ for each body literal $p_i$ of $r$;
(2.2) $\mt{definitely(\non p)} \notin W$; and
(2.3) for every $s$ either 
$\mt{lambda(q_i)} \notin W$ or
there is a $t$ with $t > s$ and $\mt{defeasibly(p'_i)} \in W$ for every body literal $p'_i$ of $t$.
By Corollary \ref{cor:strat}, $not~\mt{definitely(\non p)} \in W$ and $not~\mt{lambda(q_i)} \in W$.
Hence, using this fact and (2.1), 
$P$ contains a version of \ref{defeasibly2} instantiated by $r$ with the entire body satisfied in $W$,
except perhaps for $\mt{not\ overruled}(p)$.
Furthermore, for every instantiated version of \ref{overruled} by a rule $s$ for $\non p$ in $P$
either $\mt{lambda(q_i)}$ is not satisfied for some body literal $q_i$ of $s$, or
there is an instantiated version of \ref{defeated} by $t$ such that the body of this version is satisfied in $W$.
In the former case, $\mt{lambda(q_i)}$ evaluates to $\false$ in $W$.
In the later case, since $W$ is a model of $P$, $\mt{defeated}(s, \non p) \in W$
Thus, in either case, the body of the version of  \ref{overruled} instantiated by $s$ evaluates to $\false$ in $W$.
It follows that $\mt{not\ overruled}(p)$ evaluates to $\true$ in $W$.
Hence the body of the version of \ref{defeasibly2} is satisfied in $W$ and consequently 
$\mt{defeasibly(p)}$ must be in $W$.  
Hence $+\pl p$ is in $X$ (by the definition of $X$).

Since this argument applies for any literal $p$, $X$ is $\pl$-deductively closed.
By Theorem \ref{thm:correct_deflam}, $X$ is deductively closed.
Hence $X \supseteq P_{\pl} \cup P_{\lambda} \cup P_{\Delta}$,
which is the smallest deductively closed set.
This establishes condition \ref{pt:psi} of Lemma \ref{lemma:corr}.

We define a function $\Gamma$ from sets of tagged literals to 3-valued interpretations. 
For any set $Z$ of tagged literals,
let $Y_Z = \{  \mt{definitely(q)} ~|~+\Delta q \in Z \} \cup
\{ \mt{lambda(q)}~|~  +\lambda q \in Z \}  \cup
\{  \mt{defeasibly(q)}~|~ +\pl q \in Z \}$
be the set of corresponding logic programming atoms.
Let $\W_P^1(I) = \W_P(I) \cup I$.
We define $\Gamma(Z) = \lfp(\W_P^1, Y_Z )$,
the least fixedpoint of $\W_P^1$ containing $Y_Z$.
Notice that this is well-defined, since $\W_P^1(f, Y)$ is a monotonic function on the sub-complete lattice
of supersets of $Y$ when $f$ is a monotonic function.
Furthermore, $\Gamma$ is monotonic, since $Z \subseteq Z'$ implies $Y_Z \subseteq Y_{Z'}$
and $\lfp(\W_P^1, Y )$ is monotonic in $Y$.
In addition,
$\W_P(\Gamma(Z)) \subseteq \Gamma(Z)$ since $\Gamma(Z)$ is a fixedpoint of $\W_P^1$.

Now, let $Z = P_{\pl} \cup P_{\lambda} \cup P_{\Delta}$ be the union of the three closures
and $U = \Gamma(Z)$.
Then $\W_P(U) \subseteq U$, that is, $U$ is a prefixedpoint of $\W_P$.
Hence $W \subseteq U = \Gamma(Z)$,
that is, $X_1 \subseteq \Gamma(X_2)$.
This establishes condition \ref{pt:gamma} of Lemma \ref{lemma:corr}.

To apply Lemma \ref{lemma:corr} we must establish that $\Psi(\Gamma(X_2)) = X_2$,
where $X_2 = P_{\pl} \cup P_{\lambda} \cup P_{\Delta}$,
but we work at a greater level of generality.
By definition of $\Gamma$, $\Gamma(Z) \supseteq Y_Z$, for any $Z$.
Hence $\Psi(\Gamma(Z)) \supseteq \Psi(Y_Z) = Z$.
For the other direction,
we know that equality holds for $\Delta$ and $\lambda$ conclusions, by Theorem~\ref{thm:correct_deflam}.
Thus we focus on $\pl$ conclusions.
We define $(\W_P^1)^{n}(Y_Z) = Y_Z$ if $n \leq 0$.

Suppose, for some deductively-closed set $Z$ of tagged literals and some literal $p$, 
$+\pl\, p \in \Psi(\Gamma(Z))\backslash Z$.
Then $\mt{defeasibly}(p) \in \Gamma(Z) \backslash Y_Z$.
Let $q$ be  one of the first such literals generated by $\W_P^1$.
That is, $\mt{defeasibly}(q) \in (\W_P^1)^{n+1}(Y_Z)$ and $\Psi((\W_P^1)^{n}(Y_Z)) = Z$.
Hence there is a strict or defeasible rule $q_1, \ldots, q_k \ARROW q$ of $D$ such that
$not~\mt{definitely}(\non q) \in (\W_P^1)^{n}(Y_Z)$,
$\{ \mt{defeasibly}(q_1),$ $\ldots, \mt{defeasibly}(q_n) \} \subseteq (\W_P^1)^{n}(Y_Z)$, and
$not~\mt{overruled}(q) \in (\W_P^1)^{n}(Y_Z)$,
since the rule in $P$ must be derived from \ref{defeasibly2}.
By Theorem~\ref{thm:correct_deflam} and Corollary~\ref{cor:strat}
we must have $+\Delta \non q \notin P_\Delta$.
Because we chose $q$ to be (one of) the first literals to be derived,
$\mt{defeasibly}(q_i) \in Y_Z$
and hence $+\pl q_i \in Z$, for $i=1,\ldots,k$.
Because all rules for $\mt{overruled}$ are derived from \ref{overruled},
for every rule $s$ for $\non q$ in $D$, say $p_1,\ldots,p_m \ARROW \non q$,
either (1) $\mt{lambda}(p_i) \notin (\W_P^1)^{n-1}(Y_Z)$, for some $i$,
or (2) $\mt{defeated}(s, q) \in (\W_P^1)^{n-1}(Y_Z)$.
If (1) then, by Theorem~\ref{thm:correct_deflam} and Corollary~\ref{cor:strat},
$+\lambda p_i \notin P_\lambda$, for some $i$.
If (2) then for some strict or defeasible rule $t$ in $D$ of the form $q'_1,\ldots,q'_h \ARROW q$,
$t > s$ and $\mt{defeasibly}(q'_i) \in (\W_P^1)^{n-2}(Y_Z)$, for $i=1,\ldots,h$
and, hence, $+\pl q'_i \in Z$, for $i=1,\ldots,h$.

In summary,
there is a strict or defeasible rule $q_1, \ldots, q_k \ARROW q$ of $D$ where:
$+\Delta \non q \notin P_\Delta$,
$+\pl q_i \in Z$, for $i=1,\ldots,k$,
and for every rule $s$ for $\non q$ in $D$, say $p_1,\ldots,p_m \ARROW \non q$,
either $+\lambda p_i \notin P_\lambda$, for some $i$, or
there is a strict or defeasible rule $t$ in $D$ of the form $q'_1,\ldots,q'_h \ARROW q$
with $t > s$ and $+\pl q'_i \in Z$, for $i=1,\ldots,h$.

Thus, by the $\pl$ inference rule, and because $Z$ is deductively closed for $D$, $+\pl q \in Z$.
This contradicts our initial supposition that $+\pl\, q \in \Psi(\Gamma(Z))\backslash Z$.
Hence there is no such $q$, and we must have $\Psi(\Gamma(Z)) = Z$.
In particular, $X_2 = P_{\pl} \cup P_{\lambda} \cup P_{\Delta}$ is deductively closed,
so $\Psi(\Gamma(X_2)) = X_2$.
This establishes condition \ref{pt:eq} of Lemma \ref{lemma:corr}.

Hence, by Lemma~\ref{lemma:corr},
$P_{\pl} \cup P_{\lambda} \cup P_{\Delta} = \Psi(W)$,
where $W$ is the well-founded model of $P_D$.
In particular,
$D \vdash +\pl q$ iff  $P_D \models_{WF}  \mt{defeasibly}(q)$.
The first part
then follows by Proposition~\ref{prop:trans1}.

The second part is established in a similar manner, using $\pl^*$ and $P_D^*$
instead of $\pl$ and $P_D$.
The argument is slightly simpler, in line with the slightly simpler inference rule of $\pl^*$
and the slightly simpler $P_D^*$.

\end{proof}
}

We can now show the correctness of computing with predicates such as $\mt{defeasibly\_q}$.

\begin{theorem}    \label{thm:correct3}
Let $D$ be a defeasible theory and $S_D$ be the transformed program.
Let $q(\vec{a})$ be a ground literal.
$q$ has the form $p$ or $\neg p$ in $D$, but
$\mt{p}$ or $\mt{not\_p}$ in the transformed metaprogram $S_D$.
\begin{itemize}
\item
$D \vdash +\Delta q(\vec{a})$  iff  $S_D \models_{WF} \mt{definitely\_q}(\vec{a})$
\item
$D \vdash +\lambda q(\vec{a})$ \,\  iff  $S_D \models_{WF}  \mt{lambda\_q}(\vec{a})$
\item
$D \vdash +\pl q(\vec{a})$ iff  $S_D \models_{WF}  \mt{defeasibly\_q}(\vec{a})$
\item
$D \vdash +\pl^* q(\vec{a})$ iff  $S^*_D \models_{WF}  \mt{defeasibly\_q}(\vec{a})$
\end{itemize}
\end{theorem}
\skipit{
\begin{proof}
By Theorems \ref{thm:correct_deflam} and \ref{thm:correct_defs}, provability from $D$ is represented by 
inference from $\M_{\pl}(D)$ (or $\M_{\pl^*}(D)$) under the well-founded semantics.
By Theorem \ref{thm:trans_all}, such inferences are equivalent to the inferences
from $S_D$ stated in the statement of this theorem.
\end{proof}
}

These results establish that the metaprograms $\M_{\pl}$ and $\M_{\pl^*}$
correctly reflect the proof-theoretic definitions of Section \ref{sec:sdl}.
However, the metaprograms are able to accommodate non-propositional defeasible theories,
and easily extend to handle constraints, which are problematic for the proof theory
when the constraint domain is infinite.
There is a strong case that metaprogram formulations should be considered the canonical
definitions for defeasible logics.

\section{Properties of the Compiled Program}   \label{sec:properties}

Some syntactic properties of $\M_{\pl}$ and $\M_{\pl^*}$ were already established in
Section \ref{sec:metaprogram}.
However, $S_D$ has these and further properties that are important to its implementation.
We now establish these properties.

\begin{theorem}   \label{thm:prop_pl}
The transformed program $S_D$ for a defeasible theory $D$ over $\DL(\pl)$ is:
\begin{enumerate}
\item
a \Datalog{} program iff $D$ is function-free
\item
variable-free iff $D$ is variable-free
\item
range-restricted iff $D$ is range-restricted 
\item
safe iff $D$ is range-restricted
\item
call-consistent 
\item
stratified if $D$ is hierarchical
\item
locally stratified if $D$ is locally hierarchical
\end{enumerate}
\end{theorem}
\skipit{
\begin{proof}
1. If $D$ is function-free then the only functions in $\M(D)$ are predicates from $D$.
After the merging of tags and predicate names, and the deletion of clauses, there are no functions remaining.
Conversely, if $S_D$ is a \Datalog{} program then no term involves a function.
But all terms of $D$ appear in $S_D$.  Hence $D$ is function-free.

2.  By inspection of the final transformed program (clauses \ref{final:definitely}--\ref{final:defeated}),
the only arguments have the form $\vec{a}$ (possibly with subscript) which come from rules in $D$
or are the argument $s$ in clauses \ref{final:overruled} and \ref{final:defeated}, which is variable-free.
Thus $S_D$ is variable-free if $D$ is variable-free.
Every term in $D$ appears in $S_D$.  Thus if $S_D$ is variable-free then $D$ is variable-free.

3.
Suppose $D$ is range-restricted.
By inspection of the final program,
all heads of clauses have arguments $\vec{a}$,
all negative literals also have arguments $\vec{a}$,
and all clauses have arguments $\vec{a}$ in a positive body literal,
except for the body clauses.
Hence these clauses are all range-restricted.
Body clauses have arguments $\vec{a}$ from the head of a rule in the head of the clause
and all the arguments $\vec{a_i}$ from the body of that rule in positive literals in the clause body.
Hence, since $D$ is range-restricted, these body clauses are range-restricted.

Conversely, every rule in $D$ is reflected in a clause of the form \ref{final:lambda_body}.
Since $S_D$ is range-restriced, \ref{final:lambda_body} is range-restricted,
and hence all rules in $D$ are range-restricted.
Furthermore, all facts in $D$ are represented in unit clauses in $S_D$, and hence are variable-free.
Thus, $D$ is range-restricted.

4.
By inspection of $S_D$, it is negation-safe:
only clauses \ref{final:overruled} could cause a problem, but the argument $s$ is a constant.
By the previous part, $S_D$ is safe iff $D$ is range-restricted.

5.
This follows from Proposition \ref{prop:stratified}, and
Theorem 6.7 from \cite{trans_pm} (or Theorem 1 from \cite{unstable}).  

6.
Let $n : \Pi \rightarrow \NN$ a mapping demonstrating the hierarchicality of $D$
and define the function $m : \Pi \rightarrow \NN$ by:

Predicates of the form $\mt{definitely\_q}$ and $\mt{body}^\Delta_r$ are mapped to $0$.

Predicates of the form $\mt{lambda\_q}$ and $\mt{body}^\lambda_r$ are mapped to $1$.

Predicates of the form $\mt{defeasibly\_q}$  are mapped to $3*n(q)+5$.

Predicates of the form $\mt{overruled\_q}$  are mapped to $3*n(q)+4$.

Predicates of the form $\mt{defeated\_q}$ and $\mt{body}^d_r$ are mapped to $3*n(q)+3$.

\noindent
It is straightforward to verify that this defines a stratification of $S_D$.
For example, to verify clauses of the form \ref{body_def} satisfy the stratification condition we have
$m( \mt{defeasibly\_{q_i}} ) = 3* n(q_i) +5 \leq 3*n(q) + 2 <  3*n(q) + 3 = m( \mt{body}^d_r )$
for each strict or defeasible rule
$
\begin{array}{lrcl}
r: &   q_1(\vec{a_1}), \ldots, q_n(\vec{a_n})       & \ARROW & q(\vec{a}) \\
\end{array}
$
for $q$,
where we use $n(q_i)  < n(q)$ from the hierarchicality of $D$.

7.
Let $n : HB \rightarrow \NN$ a mapping demonstrating the local hierarchicality of $D$
and define the function $m : \Pi \rightarrow \NN$ by:

Ground atoms of the form $\mt{definitely\_q}(\vec{a})$ and $\mt{body}^\Delta_r(\vec{a})$ are mapped to $0$.

Ground atoms of the form $\mt{lambda\_q}(\vec{a})$ and $\mt{body}^\lambda_r(\vec{a})$ are mapped to $1$.

Ground atoms of the form $\mt{defeasibly\_q}(\vec{a})$  are mapped to $3*n(q(\vec{a}))+5$.

Ground atoms of the form $\mt{overruled\_q}(\vec{a})$  are mapped to $3*n(q(\vec{a}))+4$.

Ground atoms of the form $\mt{defeated\_q}(\vec{a})$ and $\mt{body}^d_r(\vec{a})$ are mapped to $3*n(q(\vec{a}))+3$.

\noindent
The proof is essentially the same as part 6.
\end{proof}
}

Of these properties, most are a reflection of properties of $D$.
However, the call-consistency of $S_D$ is a reflection of the defeasible logic.
In $\DL(\pl)$,
in the $+\pl$ inference rule, $+\pl$ appears only positively and $-\pl$ does not appear;
in $\DL(\partial)$
in the $+\partial$ inference rule, $-\partial$ appears.
It is this difference that leads to the call-consistency of $S_D$, independent of $D$.

For the logic $\DL(\pl^*)$ we have similar properties.
An important difference with the previous theorem is that
the transformed program is stratified, whether or not the defeasible theory is hierarchical.

\begin{theorem}    \label{thm:prop_pl*}
The transformed program $S^*_D$ for a defeasible theory $D$ over $\DL(\pl^*)$ is:
\begin{enumerate}
\item
a \Datalog{} program iff $D$ is function-free
\item
variable-free iff $D$ is variable-free
\item
range-restricted iff $D$ is range-restricted 
\item
safe iff $D$ is range-restricted 
\item
stratified
\end{enumerate}
\end{theorem}
\skipit{
\begin{proof}
The arguments are essentially the same as for Theorem \ref{thm:prop_pl}.
In particular, the proof of part 5 is the same as the proof of part 5 of Theorem \ref{thm:prop_pl}
(notwithstanding the different properties addressed).
In part 4, 
clauses \ref{final:defeasibly2_indiv}, \ref{final:overruled_indiv}, and \ref{final:defeated_indiv}
are negation-safe because arguments $r$ and $s$ are constants.
\end{proof}
}

As a corollary, we have that the subset of clauses in $S^*_D$ defining predicates of the form
$\mt{definitely\_q}$ and $\mt{lambda\_q}$
(and $\mt{body}^\Delta_r$ and $\mt{body}^\lambda_r$)
is stratified.
This is of interest because the same set of clauses define $\mt{definitely\_q}$ and $\mt{lambda\_q}$ in $S_D$.

\begin{corollary}   \label{cor:strat}
Let $G$ be the subset of $S_D$ and $S^*_D$ defining predicates of the form
$\mt{definitely\_q}$, $\mt{lambda\_q}$, 
$\mt{body}^\Delta_r$, and $\mt{body}^\lambda_r$.
Then $G$ is stratified.

Let $W$ be the well-founded model of $S_D$ (or $S^*_D$).
Then, for every predicate $q$ in $D$ and every $\vec{a}$,
$\mt{definitely\_q}(\vec{a}) \notin W$ iff $not~\mt{definitely\_q}(\vec{a}) \in W$,
and
$\mt{lambda\_q}(\vec{a}) \notin W$ iff $not~\mt{lambda\_q}(\vec{a}) \in W$.
\end{corollary}
\skipit{
\begin{proof}
$G$ is stratified because $S^*_D$ stratified and $G$ is a subset of $S^*_D$.
$G$ is downward-closed, and the well-founded model is total on stratified programs \cite{WF91}.
The second part then follows.
\end{proof}
}

We saw earlier that computing the $\mt{defeasibly}$ predicate in $\M_{\pl}(D)$
can be achieved by first computing $\mt{definitely}$ and $\mt{lambda}$,
and then applying Fitting's semantics to $\mt{defeasibly}$ and related predicates.
The same basic idea applies to $S_D$.

\begin{theorem}    \label{thm:Fitting}
Let $D$ be a defeasible theory
and $S_D$ be the program transformed from $\M_{\pl}(D)$.
Let $\P$ denote the predicates of $S_D$,
and $\F$ be the set of predicates  in $S_D$ of the form 
$\mt{definitely\_q}$, $\mt{body}^\Delta_r$, $\mt{lambda\_q}$, and $\mt{body}^\lambda_r$.
Let $\Q = \P \backslash \F$.
Then
\begin{itemize}
\item 
$\P$ is downward closed with floor $\F$
\item
$\Q$ has a signing $s$ where all predicates $\mt{defeasibly\_q}$ are assigned +1.
\item
For any predicate $q$ in $D$,
the predicate $\mt{defeasibly\_q}$ avoids negative unfoundedness wrt $s$.
\end{itemize}
\end{theorem}
\skipit{
\begin{proof}
It is clear, by inspection of $S_D$, that $\P$ is downward closed with floor $\F$.
Let $s$ assign all predicates $\mt{overruled\_q}$ the value $-1$,
and all other predicates in $\Q$ the value $+1$.
It is straightforward to verify that this is a signing for $\Q$.
Furthermore, any predicate $\mt{overruled\_q}$ does not depend positively on any predicate in $\Q$
(it only depends positively on predicates $\mt{body}^\lambda_r$, which are in $\F$).
Thus, trivially, every predicate in $\Q$ avoids negative unfoundedness wrt $s$ and,
in particular, the predicates $\mt{defeasibly\_q}$.
\end{proof}
}
Thus, the predicates $\mt{defeasibly\_q}$ can be computed by
first computing $\mt{definitely\_q}$, then $\mt{lambda\_q}$
under the stratified approach,
and then applying Fitting's semantics.

This same result holds for any $S^*_D$ derived from $\M_{\pl^*}(D)$,
by essentially the same proof. 
However, it is less useful because we have already established that $S^*_D$ is stratified.

\begin{corollary}
Let $D$ be a defeasible theory
and $S^*_D$ be the program transformed from $\M_{\pl^*}(D)$.
Let $\P$ denote the predicates of $S^*_D$,
and $\F$ be the set of predicates  in $S^*_D$ of the form 
$\mt{definitely\_q}$, $\mt{body}^\Delta_r$, $\mt{lambda\_q}$, and $\mt{body}^\lambda_r$.
Let $\Q = \P \backslash \F$.
Then
\begin{itemize}
\item 
$\P$ is downward closed with floor $\F$
\item
$\Q$ has a signing $s$ where all predicates $\mt{defeasibly\_q}$ are assigned +1.
\item
For any predicate $q$ in $D$,
the predicate $\mt{defeasibly\_q}$ avoids negative unfoundedness wrt $s$.
\end{itemize}
\end{corollary}

The use of the well-founded semantics to define the meaning of $\M_{\pl}(D)$
might at first appear questionable,
especially for propositional defeasible theories.
Consequences of propositional theories can be computed in time linear in the size of the theory,
while the well-founded semantics has only a quadratic upper bound.
However,
the above results show that the consequences of a propositional $D$ in $\DL(\pl)$  (or $\DL(\pl^*)$), 
computed by the well-founded semantics from $S_D$ (or $S^*_D$),
can be derived in linear time.
This is because  the subset of $S_D$ (or $S^*_D$) defining predicates $\mt{definitely\_q}$ and $\mt{lambda\_q}$
is stratified, and the stratified semantics can be computed in linear time.
Furthermore, the size of $S_D$ is linear in the size of $D$ (from Proposition \ref{prop:linear}).
And then, by Theorems \ref{thm:Fitting} and \ref{thm:comb} part \ref{thm:EQ},
it suffices to compute the remaining predicates under Fitting's semantics,
which can be done in linear time for essentially propositional programs.

More generally,
the properties identified in this section
provide a partial basis for choosing the implementation of \Datalog{}
in which to execute the transformed metaprogram,
and to adapt to available implementations.
This will be discussed in detail in the next section.

\section{Executing Compiled Theories}   \label{sec:opt}

There are three main implementation techniques for Datalog:
top-down execution with tabling, like XSB;
grounding with propositional inference, as used in ASP systems;
and bottom-up execution, using database techniques.
In addition, there are novel implementation techniques:
\emph{bddbddb} \cite{bddbddb} is based on BDDs,
and there is promise in implementations based on linear algebra \cite{linalg_datalog},
which may be able to exploit the considerable research on software and hardware accelerators
for linear algebra.
Of the three main techniques, all can be used both for query-answering and for generating all conclusions,
although their relative efficiency for each scenario is not completely clear.

The compilation of defeasible logics to \Datalog{} is motivated by
the variety of Datalog systems,
but the effectiveness of this approach is limited by the number of implementations of the
full well-founded semantics.
For example, many systems require safe programs.
Fortunately, this is not a serious barrier in implementing defeasible logics because
defeasible theories are usually range-restricted and so,
as shown in the previous section, the compiled program is safe.

However, many systems have other shortcomings.
While some implementations aim to compute the well-founded model,
others only apply to stratified programs, and others are in between.
Nevertheless, they may be sufficient for executing some defeasible theories
and/or providing an approximation of the conclusions for theories.
In this section we explore these possibilities.
We will briefly discuss the various systems and how the properties presented in the previous section
can be used to adapt to limitations of the systems.
The systems are organized by the semantics that they can compute.

\subsection{Well-Founded Semantics}

XSB \cite{XSB} is based on top-down execution with tabling, SLG-resolution.
Although SLG-resolution is complete for function-free programs \cite{SLG},
the current version of XSB does not implement answer completion (\cite{XSB2}, page 109),
one of the operations employed by SLG-resolution.
Although it appears that this is only needed in pathological examples,
the result is that, in general, XSB is sound but not complete \cite{XSB2}.\footnote{
A complicating issue is that rules such as $p(X) \Rightarrow p(X)$,
which can lead to such pathological examples,
can be used in defeasible theories to ensure that some $p$-conclusions are undefined, rather than disproved.
Thus the incompleteness of XSB might not affect this usage.
}

Early versions of Smodels \cite{Smodels_old} computed the well-founded model,
but the system is no longer at the state of the art.
DLV \cite{DLV} is primarily aimed at computing answer sets for disjunctive logic programs,
but the switch \texttt{-wf} allows the computation of the well-founded model.
On the other hand,
\textit{clingo} \cite{clingo} does not provide direct access to the well-founded model,
but it does provide an indirect way to compute the well-founded model of $S_D$.
It has the switch \texttt{-e cautious} which generates all literals true in all stable models,
and thus gives an upper bound of the well-founded model.
Furthermore, by Theorem~10 in the appendix of \cite{signings} (adapted from Theorem 5.11 of \cite{Dung95}),
for $S_D$ this gives exactly the well-founded model.
However, this is likely to be a quite inefficient way to compute the consequences of $D$.

IRIS \cite{IRIS} was designed as a platform for implementing languages such as RDFS and
description logics.
It computes the well-founded semantics,
with a choice of techniques, but development seems to have stalled in 2010.

The system of \cite{WFS_bigdata} provides a implementation of the well-founded semantics,
based on the MapReduce framework \cite{MapReduce}.
An earlier system \cite{strat_bigdata} implemented the semantics only for stratified programs.
These are more proof-of-concept implementations than production-level  systems.

All these systems can compute the consequences of a defeasible theory $D$ in $\DL(\pl)$.
Furthermore, from Theorem~\ref{thm:comb}, for most predicates only the positive
or only the negative part needs to be computed,
although it appears that this distinction is only useful for bottom-up implementations,
such as IRIS and the system of \cite{WFS_bigdata}.

\subsection{Stratified Semantics}

The stratified semantics 
(that is, the well-founded semantics computed only when the program is stratified)
is possibly the easiest form of negation to implement,
since it requires only a simple syntactic analysis and the layering of negation-free subprograms.
There are numerous systems that compute this semantics,
including
LogicBlox \cite{LogicBlox}, Souffl\'{e}  \cite{Souffle}, QL \cite{QL}, RecStep \cite{RecStep}, VLog \cite{VLog}, and Formulog \cite{Formulog}.
Furthermore, the grounders \textit{gringo} \cite{gringo} and I-DLV \cite{I-DLV},
for \textit{clingo} and DLV respectively,
will compute the well-founded model for stratified programs.
Finally, any implementation of Datalog (without negation) can be used as the basis
for an implementation of the stratified semantics, using a scripting language, for example.
However, that would come with a significant drag on performance,
compared with an integrated implementation of the stratified semantics.

Recall that we are only interested in conclusions of the form $+\Delta q(\vec{a})$ and $+\pl q(\vec{a})$, 
for literals $q(\vec{a})$,
and hence only interested in the computation of atoms of the form
$\mt{defeasibly\_q}(\vec{a})$
or 
$\mt{delta\_q}(\vec{a})$.
The latter can be computed exactly by a system supporting the stratified semantics,
while the program for the former is not, in general, stratified.
However, 
if $D$ is hierarchical then $S_D$ is stratified (Theorem \ref{thm:prop_pl}.6) and such systems can compute
exactly the consequences of $\M_{\pl}(D)$.

Even when $D$ is not hierarchical, these systems 
can provide a sound approximation to $\M_{\pl}(D)$.
As we have seen (Theorem \ref{thm:prop_pl*}), $S^*_D$ is stratified.
Consequently, these systems  can compute
the set of $\mt{defeasibly\_q}(\vec{a})$ atoms computed from $S^*_D$.
As established in \cite{sdl} (Theorem 11), $\pl^* \subset \pl$ so
this set of atoms is a sound approximation of the consequences of $\M_{\pl}(D)$.
Furthermore,  stratified fragments of $S_D$ can be used to potentially improve the sound approximation.
In some cases, it can be necessary to alternate the use of $S^*_D$ and stratified fragments of $S_D$.

\subsection{Intermediate and Ad Hoc Semantics}

Other semantics for \Datalog{} are less frequently targeted.
Locally hierarchical defeasible theories need the implementation of well-founded semantics
only for locally stratified programs (Theorem~\ref{thm:prop_pl}), which falls in between the stratified semantics and
the full well-founded semantics.
Few, if any, implementations specifically address this class.

Nevertheless, the output of the grounder \textit{gringo} \cite{gringo} can be used to infer 
 an underestimate of the well-founded model $W$ of the input program.
 From its output we can determine $a \in W$ if the fact $a$ is output, and
 $not~a \in W$ if no rule for $a$ is output;
 this provides us with an underestimate of $W$.
 This underestimate is, in fact, exact for stratified programs  \cite{Roland}.
For locally stratified programs,
 repeated application of \textit{gringo} on its output can lead to exactly the well-founded model \cite{Roland}.
Possibly {\cal I}-DLV \cite{I-DLV} can achieve the same outcome.

Theorems \ref{thm:comb} part \ref{thm:EQ} and \ref{thm:Fitting} show that 
the $\mt{defeasibly\_q}$ predicates can be computed
using a combination of the stratified and Fitting semantics.
Although only at the research stage,
the linear algebraic approach is capable of this combination,
since it handles definite clauses \cite{linalg_datalog} and Fitting's semantics \cite{linalg_3val}.

Finally,
in theory, implementations need only compute one part of each predicate not in the floor
(from Theorem~\ref{thm:comb}).
However, as mentioned earlier,
it depends very much on the implementation technique whether this property
can be exploited to improve performance.

\section{Conclusions}

In this paper we have formulated a metaprogram representation of the defeasible logic $\DL(\pl)$
and proved it correct.
We used established transformations to derive a correct compilation of defeasible theories
to \Datalog{} programs.
And, using properties of the consequent programs,
we outlined how they can be used to adapt to limitations of an underlying Datalog system.

Although we focussed on the logic $\DL(\pl)$,
the metaprogramming  and transformation approach applies to any defeasible logic.
However the design of $\DL(\pl)$, motivated by scalability,
induced structure on the resulting \Datalog{} that can simplify computation and/or made it more efficient
than the result of compilation for other defeasible logics (such as $\DL(\partial)$).
This structure also supported adaptation and approximation in response to the
limited availability of implementations of the full well-founded semantics for \Datalog{}.
It suggests that the designers of new defeasible logics should take the possibility of similar structures
into account during the design of these logics.

We also provided more evidence of the usefulness of using a metaprogram to define a defeasible logic,
rather than inference rules such as those in Sections \ref{sec:defeasible_logic} and \ref{sec:sdl}.
Those inference rules do not generalize easily to non-propositional defeasible theories over infinite domains,
whereas that is a non-issue for metaprograms.
Furthermore, such inference rules are difficult to reason about.
Metaprograms provide access to all the tools of (constraint) logic programming for reasoning about
and implementing the defeasible logics.

The original motivation for this work was to provide alternatives to the bespoke implementation of $\DL(\pl)$
in \cite{sdl}.
So, it was disappointing to realise the limited range of Datalog implementations available
that support the full well-founded semantics.
There remains much scope for implementations of the well-founded semantics
on novel architectures.

\textbf{Acknowledgements:}
This paper is dedicated to the memory of Christian Schulte.

The author has an adjunct position at Griffith University and an honorary position at UNSW.
He thanks the reviewers for their thorough reviews and comments, which helped improve the paper.

\textbf{Competing interests:} 
The author declares none.

\bibliography{sdl2}

\begin{thebibliography}{}

\bibitem[\protect\citeauthoryear{Aberger, Lamb, Tu, N{\"{o}}tzli, Olukotun, and
  R{\'{e}}}{Aberger et~al\mbox{.}}{2017}]{EmptyHeaded}
{\sc Aberger, C.~R.}, {\sc Lamb, A.}, {\sc Tu, S.}, {\sc N{\"{o}}tzli, A.},
  {\sc Olukotun, K.}, {\sc and} {\sc R{\'{e}}, C.} 2017.
\newblock Emptyheaded: {A} relational engine for graph processing.
\newblock {\em {ACM} Trans. Database Syst.\/}~{\em 42,\/}~4, 20:1--20:44.

\bibitem[\protect\citeauthoryear{Abiteboul, Hull, and Vianu}{Abiteboul
  et~al\mbox{.}}{1995}]{AHV}
{\sc Abiteboul, S.}, {\sc Hull, R.}, {\sc and} {\sc Vianu, V.} 1995.
\newblock {\em Foundations of Databases}.
\newblock Addison-Wesley.

\bibitem[\protect\citeauthoryear{Adrian, Alviano, Calimeri, Cuteri, Dodaro,
  Faber, Fusc{\`{a}}, Leone, Manna, Perri, Ricca, Veltri, and Zangari}{Adrian
  et~al\mbox{.}}{2018}]{DLV}
{\sc Adrian, W.~T.}, {\sc Alviano, M.}, {\sc Calimeri, F.}, {\sc Cuteri, B.},
  {\sc Dodaro, C.}, {\sc Faber, W.}, {\sc Fusc{\`{a}}, D.}, {\sc Leone, N.},
  {\sc Manna, M.}, {\sc Perri, S.}, {\sc Ricca, F.}, {\sc Veltri, P.}, {\sc
  and} {\sc Zangari, J.} 2018.
\newblock The {ASP} system {DLV:} advancements and applications.
\newblock {\em {KI}\/}~{\em 32,\/}~2-3, 177--179.

\bibitem[\protect\citeauthoryear{Alvaro, Marczak, Conway, Hellerstein, Maier,
  and Sears}{Alvaro et~al\mbox{.}}{2010}]{Dedalus}
{\sc Alvaro, P.}, {\sc Marczak, W.~R.}, {\sc Conway, N.}, {\sc Hellerstein,
  J.~M.}, {\sc Maier, D.}, {\sc and} {\sc Sears, R.} 2010.
\newblock Dedalus: {Datalog} in time and space.
\newblock In {\em Datalog Reloaded - First International Workshop, Datalog
  2010}, {O.~de~Moor}, {G.~Gottlob}, {T.~Furche}, {and} {A.~J. Sellers}, Eds.
  Lecture Notes in Computer Science, vol. 6702. Springer, 262--281.

\bibitem[\protect\citeauthoryear{Antoniou, Billington, Governatori, and
  Maher}{Antoniou et~al\mbox{.}}{2000}]{flexf}
{\sc Antoniou, G.}, {\sc Billington, D.}, {\sc Governatori, G.}, {\sc and} {\sc
  Maher, M.~J.} 2000.
\newblock A flexible framework for defeasible logics.
\newblock In {\em AAAI/IAAI}. AAAI Press / The MIT Press, 405--410.

\bibitem[\protect\citeauthoryear{Antoniou, Billington, Governatori, and
  Maher}{Antoniou et~al\mbox{.}}{2001}]{TOCL01}
{\sc Antoniou, G.}, {\sc Billington, D.}, {\sc Governatori, G.}, {\sc and} {\sc
  Maher, M.~J.} 2001.
\newblock Representation results for defeasible logic.
\newblock {\em ACM Trans. Comput. Log.\/}~{\em 2,\/}~2, 255--287.

\bibitem[\protect\citeauthoryear{Antoniou, Billington, Governatori, and
  Maher}{Antoniou et~al\mbox{.}}{2006}]{TPLP06}
{\sc Antoniou, G.}, {\sc Billington, D.}, {\sc Governatori, G.}, {\sc and} {\sc
  Maher, M.~J.} 2006.
\newblock Embedding defeasible logic into logic programming.
\newblock {\em Theory Pract. Log. Program.\/}~{\em 6,\/}~6, 703--735.

\bibitem[\protect\citeauthoryear{Antoniou, Billington, Governatori, Maher, and
  Rock}{Antoniou et~al\mbox{.}}{2000}]{ECAI00}
{\sc Antoniou, G.}, {\sc Billington, D.}, {\sc Governatori, G.}, {\sc Maher,
  M.~J.}, {\sc and} {\sc Rock, A.} 2000.
\newblock A family of defeasible reasoning logics and its implementation.
\newblock In {\em ECAI}, {W.~Horn}, Ed. IOS Press, 459--463.

\bibitem[\protect\citeauthoryear{Antoniou and Maher}{Antoniou and
  Maher}{2002}]{embed}
{\sc Antoniou, G.} {\sc and} {\sc Maher, M.~J.} 2002.
\newblock Embedding defeasible logic into logic programs.
\newblock In {\em Logic Programming, 18th International Conference}. 393--404.

\bibitem[\protect\citeauthoryear{Apt, Blair, and Walker}{Apt
  et~al\mbox{.}}{1988}]{ABW}
{\sc Apt, K.~R.}, {\sc Blair, H.~A.}, {\sc and} {\sc Walker, A.} 1988.
\newblock Towards a theory of declarative knowledge.
\newblock In {\em Foundations of Deductive Databases and Logic Programming.}
  Morgan Kaufmann, 89--148.

\bibitem[\protect\citeauthoryear{Apt and Bol}{Apt and Bol}{1994}]{AptBol}
{\sc Apt, K.~R.} {\sc and} {\sc Bol, R.~N.} 1994.
\newblock Logic programming and negation: A survey.
\newblock {\em J. Log. Program.\/}~{\em 19/20}, 9--71.

\bibitem[\protect\citeauthoryear{Aravindan and Dung}{Aravindan and
  Dung}{1995}]{Aravindan}
{\sc Aravindan, C.} {\sc and} {\sc Dung, P.~M.} 1995.
\newblock On the correctness of unfold/fold transformation of normal and
  extended logic programs.
\newblock {\em J. Log. Program.\/}~{\em 24,\/}~3, 201--217.

\bibitem[\protect\citeauthoryear{Aref, ten Cate, Green, Kimelfeld, Olteanu,
  Pasalic, Veldhuizen, and Washburn}{Aref et~al\mbox{.}}{2015}]{LogicBlox}
{\sc Aref, M.}, {\sc ten Cate, B.}, {\sc Green, T.~J.}, {\sc Kimelfeld, B.},
  {\sc Olteanu, D.}, {\sc Pasalic, E.}, {\sc Veldhuizen, T.~L.}, {\sc and} {\sc
  Washburn, G.} 2015.
\newblock Design and implementation of the {LogicBlox} system.
\newblock In {\em Proceedings of the 2015 {ACM} {SIGMOD} International
  Conference on Management of Data}, {T.~K. Sellis}, {S.~B. Davidson}, {and}
  {Z.~G. Ives}, Eds. {ACM}, 1371--1382.

\bibitem[\protect\citeauthoryear{Avgustinov, de~Moor, Jones, and
  Sch{\"{a}}fer}{Avgustinov et~al\mbox{.}}{2016}]{QL}
{\sc Avgustinov, P.}, {\sc de~Moor, O.}, {\sc Jones, M.~P.}, {\sc and} {\sc
  Sch{\"{a}}fer, M.} 2016.
\newblock {QL:} object-oriented queries on relational data.
\newblock In {\em 30th European Conference on Object-Oriented Programming,
  {ECOOP} 2016}, {S.~Krishnamurthi} {and} {B.~S. Lerner}, Eds. LIPIcs, vol.~56.
  Schloss Dagstuhl - Leibniz-Zentrum f{\"{u}}r Informatik, 2:1--2:25.

\bibitem[\protect\citeauthoryear{Bassiliades, Antoniou, and
  Vlahavas}{Bassiliades et~al\mbox{.}}{2006}]{DR-DEVICE}
{\sc Bassiliades, N.}, {\sc Antoniou, G.}, {\sc and} {\sc Vlahavas, I.~P.}
  2006.
\newblock A defeasible logic reasoner for the semantic web.
\newblock {\em Int. J. Semantic Web Inf. Syst.\/}~{\em 2,\/}~1, 1--41.

\bibitem[\protect\citeauthoryear{Bellomarini, Sallinger, and
  Gottlob}{Bellomarini et~al\mbox{.}}{2018}]{Vadalog}
{\sc Bellomarini, L.}, {\sc Sallinger, E.}, {\sc and} {\sc Gottlob, G.} 2018.
\newblock The {Vadalog} system: {Datalog}-based reasoning for knowledge graphs.
\newblock {\em Proc. {VLDB} Endow.\/}~{\em 11,\/}~9, 975--987.

\bibitem[\protect\citeauthoryear{Bembenek, Greenberg, and Chong}{Bembenek
  et~al\mbox{.}}{2020}]{Formulog}
{\sc Bembenek, A.}, {\sc Greenberg, M.}, {\sc and} {\sc Chong, S.} 2020.
\newblock Formulog: Datalog for smt-based static analysis.
\newblock {\em Proc. {ACM} Program. Lang.\/}~{\em 4,\/}~{OOPSLA},
  141:1--141:31.

\bibitem[\protect\citeauthoryear{Billington, Antoniou, Governatori, and
  Maher}{Billington et~al\mbox{.}}{2010}]{TOCL10}
{\sc Billington, D.}, {\sc Antoniou, G.}, {\sc Governatori, G.}, {\sc and} {\sc
  Maher, M.~J.} 2010.
\newblock An inclusion theorem for defeasible logics.
\newblock {\em ACM Trans. Comput. Log.\/}~{\em 12,\/}~1, 6.

\bibitem[\protect\citeauthoryear{Billington and Rock}{Billington and
  Rock}{2001}]{phobos}
{\sc Billington, D.} {\sc and} {\sc Rock, A.} 2001.
\newblock Propositional plausible logic: Introduction and implementation.
\newblock {\em Studia Logica\/}~{\em 67,\/}~2, 243--269.

\bibitem[\protect\citeauthoryear{Bishop and Fischer}{Bishop and
  Fischer}{2008}]{IRIS}
{\sc Bishop, B.} {\sc and} {\sc Fischer, F.} 2008.
\newblock {IRIS} - integrated rule inference system.
\newblock In {\em Proc. Workshop on Advancing Reasoning on the Web: Scalability
  and Commonsense}. Vol. 350. CEUR Workshop Proceedings.

\bibitem[\protect\citeauthoryear{Bossi, Cocco, and Etalle}{Bossi
  et~al\mbox{.}}{1992}]{BossiCE92}
{\sc Bossi, A.}, {\sc Cocco, N.}, {\sc and} {\sc Etalle, S.} 1992.
\newblock Transforming normal programs by replacement.
\newblock In {\em Proc. Meta-Programming in Logic, 3rd International Workshop}.
  Lecture Notes in Computer Science, vol. 649. Springer, 265--279.

\bibitem[\protect\citeauthoryear{Brass and Stephan}{Brass and
  Stephan}{2017}]{push}
{\sc Brass, S.} {\sc and} {\sc Stephan, H.} 2017.
\newblock Pipelined bottom-up evaluation of {Datalog} programs: The push
  method.
\newblock In {\em Perspectives of System Informatics - 11th International
  Andrei P. Ershov Informatics Conference, {PSI} 2017}, {A.~K. Petrenko} {and}
  {A.~Voronkov}, Eds. Lecture Notes in Computer Science, vol. 10742. Springer,
  43--58.

\bibitem[\protect\citeauthoryear{Brewka}{Brewka}{1996}]{Brewka96}
{\sc Brewka, G.} 1996.
\newblock Well-founded semantics for extended logic programs with dynamic
  preferences.
\newblock {\em J. Artif. Intell. Res.\/}~{\em 4}, 19--36.

\bibitem[\protect\citeauthoryear{Brewka}{Brewka}{2001}]{Brewka01}
{\sc Brewka, G.} 2001.
\newblock On the relationship between defeasible logic and well-founded
  semantics.
\newblock In {\em LPNMR}. 121--132.

\bibitem[\protect\citeauthoryear{Calimeri, Fusc{\`{a}}, Perri, and
  Zangari}{Calimeri et~al\mbox{.}}{2017}]{I-DLV}
{\sc Calimeri, F.}, {\sc Fusc{\`{a}}, D.}, {\sc Perri, S.}, {\sc and} {\sc
  Zangari, J.} 2017.
\newblock {I-DLV:} the new intelligent grounder of {DLV}.
\newblock {\em Intelligenza Artificiale\/}~{\em 11,\/}~1, 5--20.

\bibitem[\protect\citeauthoryear{Carral, Dragoste, Gonz{\'{a}}lez, Jacobs,
  Kr{\"{o}}tzsch, and Urbani}{Carral et~al\mbox{.}}{2019}]{VLog}
{\sc Carral, D.}, {\sc Dragoste, I.}, {\sc Gonz{\'{a}}lez, L.}, {\sc Jacobs, C.
  J.~H.}, {\sc Kr{\"{o}}tzsch, M.}, {\sc and} {\sc Urbani, J.} 2019.
\newblock Vlog: {A} rule engine for knowledge graphs.
\newblock In {\em The Semantic Web - {ISWC} 2019 - 18th International Semantic
  Web Conference, Part {II}}, {C.~Ghidini}, {O.~Hartig}, {M.~Maleshkova},
  {V.~Sv{\'{a}}tek}, {I.~F. Cruz}, {A.~Hogan}, {J.~Song},
  {M.~Lefran{\c{c}}ois}, {and} {F.~Gandon}, Eds. Lecture Notes in Computer
  Science, vol. 11779. Springer, 19--35.

\bibitem[\protect\citeauthoryear{Cat, Bogaerts, Bruynooghe, Janssens, and
  Denecker}{Cat et~al\mbox{.}}{2018}]{IDP}
{\sc Cat, B.~D.}, {\sc Bogaerts, B.}, {\sc Bruynooghe, M.}, {\sc Janssens, G.},
  {\sc and} {\sc Denecker, M.} 2018.
\newblock Predicate logic as a modeling language: the {IDP} system.
\newblock In {\em Declarative Logic Programming: Theory, Systems, and
  Applications}. 279--323.

\bibitem[\protect\citeauthoryear{Chen and Warren}{Chen and Warren}{1996}]{SLG}
{\sc Chen, W.} {\sc and} {\sc Warren, D.~S.} 1996.
\newblock Tabled evaluation with delaying for general logic programs.
\newblock {\em J. {ACM}\/}~{\em 43,\/}~1, 20--74.

\bibitem[\protect\citeauthoryear{Chin, von Dincklage, Ercegovac, Hawkins,
  Miller, Och, Olston, and Pereira}{Chin et~al\mbox{.}}{2015}]{Yedalog}
{\sc Chin, B.}, {\sc von Dincklage, D.}, {\sc Ercegovac, V.}, {\sc Hawkins,
  P.}, {\sc Miller, M.~S.}, {\sc Och, F.~J.}, {\sc Olston, C.}, {\sc and} {\sc
  Pereira, F.} 2015.
\newblock Yedalog: Exploring knowledge at scale.
\newblock In {\em 1st Summit on Advances in Programming Languages, {SNAPL}
  2015, May 3-6, 2015, Asilomar, California, {USA}}, {T.~Ball},
  {R.~Bod{\'{\i}}k}, {S.~Krishnamurthi}, {B.~S. Lerner}, {and} {G.~Morrisett},
  Eds. LIPIcs, vol.~32. Schloss Dagstuhl - Leibniz-Zentrum f{\"{u}}r
  Informatik, 63--78.

\bibitem[\protect\citeauthoryear{{Cognitect}}{{Cognitect}}{}]{Datomic}
{\sc {Cognitect}}.
\newblock What is datomic cloud?
\newblock {https://docs.datomic.com/cloud/index.html}.
\newblock Accessed 8/2/2020.

\bibitem[\protect\citeauthoryear{Condie, Das, Interlandi, Shkapsky, Yang, and
  Zaniolo}{Condie et~al\mbox{.}}{2018}]{BigDatalog}
{\sc Condie, T.}, {\sc Das, A.}, {\sc Interlandi, M.}, {\sc Shkapsky, A.}, {\sc
  Yang, M.}, {\sc and} {\sc Zaniolo, C.} 2018.
\newblock Scaling-up reasoning and advanced analytics on bigdata.
\newblock {\em Theory Pract. Log. Program.\/}~{\em 18,\/}~5-6, 806--845.

\bibitem[\protect\citeauthoryear{Costa, Rocha, and Damas}{Costa
  et~al\mbox{.}}{2012}]{YAP}
{\sc Costa, V.~S.}, {\sc Rocha, R.}, {\sc and} {\sc Damas, L.} 2012.
\newblock The {YAP Prolog} system.
\newblock {\em Theory Pract. Log. Program.\/}~{\em 12,\/}~1-2, 5--34.

\bibitem[\protect\citeauthoryear{Covington, Nute, and Vellino}{Covington
  et~al\mbox{.}}{1997}]{Nute_subsumption}
{\sc Covington, M.}, {\sc Nute, D.}, {\sc and} {\sc Vellino, A.} 1997.
\newblock {\em Prolog Programming in Depth}.
\newblock Prentice-Hall.

\bibitem[\protect\citeauthoryear{Dean and Ghemawat}{Dean and
  Ghemawat}{2004}]{MapReduce}
{\sc Dean, J.} {\sc and} {\sc Ghemawat, S.} 2004.
\newblock {M}ap{R}educe: simplified data processing on large clusters.
\newblock In {\em 6th Symposium on Operating System Design and Implementation
  {(OSDI)}}. USENIX Association, Berkeley, CA, USA, 10--10.

\bibitem[\protect\citeauthoryear{Dung}{Dung}{1995}]{Dung95}
{\sc Dung, P.~M.} 1995.
\newblock On the acceptability of arguments and its fundamental role in
  nonmonotonic reasoning, logic programming and n-person games.
\newblock {\em Artif. Intell.\/}~{\em 77,\/}~2, 321--358.

\bibitem[\protect\citeauthoryear{Eisner and Filardo}{Eisner and
  Filardo}{2010}]{Dyna}
{\sc Eisner, J.} {\sc and} {\sc Filardo, N.~W.} 2010.
\newblock Dyna: Extending {Datalog} for modern {AI}.
\newblock In {\em Datalog Reloaded}. 181--220.

\bibitem[\protect\citeauthoryear{Eiter, Leone, and Sacc{\`{a}}}{Eiter
  et~al\mbox{.}}{1997}]{Lstable}
{\sc Eiter, T.}, {\sc Leone, N.}, {\sc and} {\sc Sacc{\`{a}}, D.} 1997.
\newblock On the partial semantics for disjunctive deductive databases.
\newblock {\em Ann. Math. Artif. Intell.\/}~{\em 19,\/}~1-2, 59--96.

\bibitem[\protect\citeauthoryear{Fan, Zhu, Zhang, Albarghouthi, Koutris, and
  Patel}{Fan et~al\mbox{.}}{2019}]{RecStep}
{\sc Fan, Z.}, {\sc Zhu, J.}, {\sc Zhang, Z.}, {\sc Albarghouthi, A.}, {\sc
  Koutris, P.}, {\sc and} {\sc Patel, J.~M.} 2019.
\newblock Scaling-up in-memory {Datalog} processing: Observations and
  techniques.
\newblock {\em {PVLDB}\/}~{\em 12,\/}~6, 695--708.

\bibitem[\protect\citeauthoryear{Filardo}{Filardo}{2017}]{Filardo_thesis}
{\sc Filardo, N.~W.} 2017.
\newblock Dyna 2: Towards a general weighted logic language.
\newblock Ph.D. thesis, Johns Hopkins University.

\bibitem[\protect\citeauthoryear{Fitting}{Fitting}{1985}]{Fitting}
{\sc Fitting, M.} 1985.
\newblock A {K}ripke-{K}leene semantics for logic programs.
\newblock {\em J. Log. Program.\/}~{\em 2,\/}~4, 295--312.

\bibitem[\protect\citeauthoryear{Gandhe, Finin, and Grosof}{Gandhe
  et~al\mbox{.}}{2002}]{SweetJess}
{\sc Gandhe, M.}, {\sc Finin, T.}, {\sc and} {\sc Grosof, B.} 2002.
\newblock {SweetJess: Translating DamlRuleML to Jess}.
\newblock In {\em International Workshop on Rule Markup Languages for Business
  Rules on the Semantic Web in conjunction with ISWC2002}. Sardinia, Italy.

\bibitem[\protect\citeauthoryear{Gebser, Kaminski, Kaufmann, and Schaub}{Gebser
  et~al\mbox{.}}{2019}]{clingo}
{\sc Gebser, M.}, {\sc Kaminski, R.}, {\sc Kaufmann, B.}, {\sc and} {\sc
  Schaub, T.} 2019.
\newblock Multi-shot {ASP} solving with clingo.
\newblock {\em Theory Pract. Log. Program.\/}~{\em 19,\/}~1, 27--82.

\bibitem[\protect\citeauthoryear{Gebser, Kaminski, K{\"{o}}nig, and
  Schaub}{Gebser et~al\mbox{.}}{2011}]{gringo}
{\sc Gebser, M.}, {\sc Kaminski, R.}, {\sc K{\"{o}}nig, A.}, {\sc and} {\sc
  Schaub, T.} 2011.
\newblock Advances in \emph{gringo} series 3.
\newblock In {\em Logic Programming and Nonmonotonic Reasoning - 11th
  International Conference, {LPNMR}}. 345--351.

\bibitem[\protect\citeauthoryear{Gelfond and Lifschitz}{Gelfond and
  Lifschitz}{1988}]{stable}
{\sc Gelfond, M.} {\sc and} {\sc Lifschitz, V.} 1988.
\newblock The stable model semantics for logic programming.
\newblock In {\em Proc. JICSLP}. 1070--1080.

\bibitem[\protect\citeauthoryear{Governatori and Maher}{Governatori and
  Maher}{2017}]{GM17}
{\sc Governatori, G.} {\sc and} {\sc Maher, M.~J.} 2017.
\newblock Annotated defeasible logic.
\newblock {\em Theory Pract. Log. Program.\/}~{\em 17,\/}~5-6, 819--836.

\bibitem[\protect\citeauthoryear{Governatori, Maher, Antoniou, and
  Billington}{Governatori et~al\mbox{.}}{2004}]{JLC04}
{\sc Governatori, G.}, {\sc Maher, M.~J.}, {\sc Antoniou, G.}, {\sc and} {\sc
  Billington, D.} 2004.
\newblock Argumentation semantics for defeasible logic.
\newblock {\em J. Log. Comput.\/}~{\em 14,\/}~5, 675--702.

\bibitem[\protect\citeauthoryear{Grosof and Kifer}{Grosof and
  Kifer}{2013}]{Rulelog}
{\sc Grosof, B.} {\sc and} {\sc Kifer, M.} 2013.
\newblock Rulelog: Syntax and semantics.
\newblock http://ruleml.org/rif/rulelog/spec/Rulelog.html.
\newblock Accessed: April 2015.

\bibitem[\protect\citeauthoryear{Grosof}{Grosof}{1997}]{Grosof97}
{\sc Grosof, B.~N.} 1997.
\newblock Prioritized conflict handling for logic programs.
\newblock In {\em ILPS}. 197--211.

\bibitem[\protect\citeauthoryear{Hoder, Bj{\o}rner, and de~Moura}{Hoder
  et~al\mbox{.}}{2011}]{muZ}
{\sc Hoder, K.}, {\sc Bj{\o}rner, N.}, {\sc and} {\sc de~Moura, L.~M.} 2011.
\newblock \emph{{\(\mu\)}Z}- an efficient engine for fixed points with
  constraints.
\newblock In {\em Computer Aided Verification - 23rd International Conference,
  {CAV}}. 457--462.

\bibitem[\protect\citeauthoryear{Jaffar and Maher}{Jaffar and
  Maher}{1994}]{JM94}
{\sc Jaffar, J.} {\sc and} {\sc Maher, M.~J.} 1994.
\newblock Constraint logic programming: {A} survey.
\newblock {\em The Journal of Logic Programming\/}~{\em 19 \& 20}, 503--582.

\bibitem[\protect\citeauthoryear{Jordan, Scholz, and Subotic}{Jordan
  et~al\mbox{.}}{2016}]{Souffle}
{\sc Jordan, H.}, {\sc Scholz, B.}, {\sc and} {\sc Subotic, P.} 2016.
\newblock Souffl{\'{e}}: On synthesis of program analyzers.
\newblock In {\em Computer Aided Verification}. 422--430.

\bibitem[\protect\citeauthoryear{Kakas and Mancarella}{Kakas and
  Mancarella}{1991}]{stable_theory}
{\sc Kakas, A.~C.} {\sc and} {\sc Mancarella, P.} 1991.
\newblock Negation as stable hypotheses.
\newblock In {\em Logic Programming and Non-monotonic Reasoning, Proceedings of
  the First International Workshop}. 275--288.

\bibitem[\protect\citeauthoryear{Kaminski}{Kaminski}{2020}]{Roland}
{\sc Kaminski, R.} 2020.
\newblock personal communication.

\bibitem[\protect\citeauthoryear{Kifer, Yang, Wan, and Zhao}{Kifer
  et~al\mbox{.}}{2018}]{Flora2}
{\sc Kifer, M.}, {\sc Yang, G.}, {\sc Wan, H.}, {\sc and} {\sc Zhao, C.} 2018.
\newblock Flora-2 documentation.
\newblock {http://flora.sourceforge.net/documentation.html}.
\newblock Accessed 14/4/2020.

\bibitem[\protect\citeauthoryear{Kunen}{Kunen}{1987}]{Kunen87}
{\sc Kunen, K.} 1987.
\newblock Negation in logic programming.
\newblock {\em J. Log. Program.\/}~{\em 4,\/}~4, 289--308.

\bibitem[\protect\citeauthoryear{Kunen}{Kunen}{1989}]{Kunen89}
{\sc Kunen, K.} 1989.
\newblock Signed data dependencies in logic programs.
\newblock {\em J. Log. Program.\/}~{\em 7,\/}~3, 231--245.

\bibitem[\protect\citeauthoryear{Lam and Governatori}{Lam and
  Governatori}{2009}]{Spindle}
{\sc Lam, H.} {\sc and} {\sc Governatori, G.} 2009.
\newblock The making of {SPINdle}.
\newblock In {\em Rule Interchange and Applications, International Symposium,
  RuleML, Proceedings}. 315--322.

\bibitem[\protect\citeauthoryear{Madsen and Lhot{\'{a}}k}{Madsen and
  Lhot{\'{a}}k}{2020}]{Ldat}
{\sc Madsen, M.} {\sc and} {\sc Lhot{\'{a}}k, O.} 2020.
\newblock Fixpoints for the masses: programming with first-class datalog
  constraints.
\newblock {\em Proc. {ACM} Program. Lang.\/}~{\em 4,\/}~{OOPSLA},
  125:1--125:28.

\bibitem[\protect\citeauthoryear{Madsen, Yee, and Lhot{\'{a}}k}{Madsen
  et~al\mbox{.}}{2016}]{Flix}
{\sc Madsen, M.}, {\sc Yee, M.}, {\sc and} {\sc Lhot{\'{a}}k, O.} 2016.
\newblock From {Datalog} to {Flix}: a declarative language for fixed points on
  lattices.
\newblock In {\em Proceedings of the 37th {ACM} {SIGPLAN} Conference on
  Programming Language Design and Implementation, {PLDI} 2016}, {C.~Krintz}
  {and} {E.~Berger}, Eds. {ACM}, 194--208.

\bibitem[\protect\citeauthoryear{Maher}{Maher}{1988}]{trans}
{\sc Maher, M.~J.} 1988.
\newblock Correctness of a logic program transformation system.
\newblock Tech. rep., IBM T.J. Watson Research Center.

\bibitem[\protect\citeauthoryear{Maher}{Maher}{1990}]{unstable}
{\sc Maher, M.~J.} 1990.
\newblock Reasoning about stable models (and other unstable semantics).
\newblock Tech. rep.

\bibitem[\protect\citeauthoryear{Maher}{Maher}{1993}]{trans_pm}
{\sc Maher, M.~J.} 1993.
\newblock A tranformation system for deductive databases modules with perfect
  model semantics.
\newblock {\em Theor. Comput. Sci.\/}~{\em 110,\/}~2, 377--403.

\bibitem[\protect\citeauthoryear{Maher}{Maher}{2002}]{Maher02}
{\sc Maher, M.~J.} 2002.
\newblock A model-theoretic semantics for defeasible logic.
\newblock In {\em Paraconsistent Computational Logic}, {H.~Decker},
  {J.~Villadsen}, {and} {T.~Waragai}, Eds. Datalogiske Skrifter, vol.~95.
  Roskilde University, Roskilde, Denmark, 67--80.

\bibitem[\protect\citeauthoryear{Maher}{Maher}{2013}]{Maher13b}
{\sc Maher, M.~J.} 2013.
\newblock Relative expressiveness of well-founded defeasible logics.
\newblock In {\em Proc. Australasian Joint Conf. on Artificial Intelligence}.

\bibitem[\protect\citeauthoryear{Maher}{Maher}{2014}]{Maher14}
{\sc Maher, M.~J.} 2014.
\newblock Comparing defeasible logics.
\newblock In {\em 21st European Conference on Artificial Intelligence}.
  585--590.

\bibitem[\protect\citeauthoryear{Maher}{Maher}{2017}]{cdr}
{\sc Maher, M.~J.} 2017.
\newblock Relating concrete defeasible reasoning formalisms and abstract
  argumentation.
\newblock {\em Fundam. Inform.\/}~{\em 155,\/}~3, 233--260.

\bibitem[\protect\citeauthoryear{Maher}{Maher}{2021}]{signings}
{\sc Maher, M.~J.} 2021.
\newblock On signings and the well-founded semantics.
\newblock {\em Theory Pract. Log. Program.\/}~accepted for publication.

\bibitem[\protect\citeauthoryear{Maher and Governatori}{Maher and
  Governatori}{1999}]{MG99}
{\sc Maher, M.~J.} {\sc and} {\sc Governatori, G.} 1999.
\newblock A semantic decomposition of defeasible logics.
\newblock In {\em AAAI/IAAI}. AAAI Press, 299--305.

\bibitem[\protect\citeauthoryear{Maher, Rock, Antoniou, Billington, and
  Miller}{Maher et~al\mbox{.}}{2001}]{MRABM}
{\sc Maher, M.~J.}, {\sc Rock, A.}, {\sc Antoniou, G.}, {\sc Billington, D.},
  {\sc and} {\sc Miller, T.} 2001.
\newblock Efficient defeasible reasoning systems.
\newblock {\em International Journal on Artificial Intelligence Tools\/}~{\em
  10,\/}~4, 483--501.

\bibitem[\protect\citeauthoryear{Maher, Tachmazidis, Antoniou, Wade, and
  Cheng}{Maher et~al\mbox{.}}{2020}]{sdl}
{\sc Maher, M.~J.}, {\sc Tachmazidis, I.}, {\sc Antoniou, G.}, {\sc Wade, S.},
  {\sc and} {\sc Cheng, L.} 2020.
\newblock Rethinking defeasible reasoning: A scalable approach.
\newblock {\em Theory Pract. Log. Program.\/}~{\em 20,\/}~4, 552--586.

\bibitem[\protect\citeauthoryear{Maier}{Maier}{2013}]{Maier13}
{\sc Maier, F.} 2013.
\newblock Interdefinability of defeasible logic and logic programming under the
  well-founded semantics.
\newblock {\em Theory Pract. Log. Program.\/}~{\em 13,\/}~1, 107--142.

\bibitem[\protect\citeauthoryear{Maier and Nute}{Maier and Nute}{2006}]{MN06}
{\sc Maier, F.} {\sc and} {\sc Nute, D.} 2006.
\newblock Ambiguity propagating defeasible logic and the well-founded
  semantics.
\newblock In {\em JELIA}, {M.~Fisher}, {W.~van~der Hoek}, {B.~Konev}, {and}
  {A.~Lisitsa}, Eds. Lecture Notes in Computer Science, vol. 4160. Springer,
  306--318.

\bibitem[\protect\citeauthoryear{Maier and Nute}{Maier and Nute}{2010}]{MN10}
{\sc Maier, F.} {\sc and} {\sc Nute, D.} 2010.
\newblock Well-founded semantics for defeasible logic.
\newblock {\em Synthese\/}~{\em 176,\/}~2, 243--274.

\bibitem[\protect\citeauthoryear{Martinez{-}Angeles, de~Castro~Dutra, Costa,
  and Buenabad{-}Ch{\'{a}}vez}{Martinez{-}Angeles
  et~al\mbox{.}}{2013}]{GPU_datalog}
{\sc Martinez{-}Angeles, C.~A.}, {\sc de~Castro~Dutra, I.}, {\sc Costa, V.~S.},
  {\sc and} {\sc Buenabad{-}Ch{\'{a}}vez, J.} 2013.
\newblock A {Datalog} engine for {GPUs}.
\newblock In {\em Declarative Programming and Knowledge Management}. 152--168.

\bibitem[\protect\citeauthoryear{Marz}{Marz}{2013}]{Cascalog}
{\sc Marz, N.} 2013.
\newblock Cascalog.
\newblock https://www.cascading.org/projects/cascalog/.
\newblock Accessed: April 2020.

\bibitem[\protect\citeauthoryear{Nguyen, Sakama, Sato, and Inoue}{Nguyen
  et~al\mbox{.}}{2018}]{lin_alg}
{\sc Nguyen, H.~D.}, {\sc Sakama, C.}, {\sc Sato, T.}, {\sc and} {\sc Inoue,
  K.} 2018.
\newblock Computing logic programming semantics in linear algebra.
\newblock In {\em Multi-disciplinary Trends in Artificial Intelligence - 12th
  International Conference, {MIWAI}}. 32--48.

\bibitem[\protect\citeauthoryear{Niemel{\"{a}} and Simons}{Niemel{\"{a}} and
  Simons}{1997}]{Smodels_old}
{\sc Niemel{\"{a}}, I.} {\sc and} {\sc Simons, P.} 1997.
\newblock Smodels - an implementation of the stable model and well-founded
  semantics for normal {LP}.
\newblock In {\em Logic Programming and Nonmonotonic Reasoning, 4th
  International Conference, {LPNMR'97}}, {J.~Dix}, {U.~Furbach}, {and}
  {A.~Nerode}, Eds. Lecture Notes in Computer Science, vol. 1265. Springer,
  421--430.

\bibitem[\protect\citeauthoryear{Nute}{Nute}{1993}]{d-Prolog}
{\sc Nute, D.} 1993.
\newblock Defeasible {Prolog}.
\newblock In {\em Proc. {AAAI Fall Symposium} on Automated Deduction in
  Nonstandard Logics}. 105--112.

\bibitem[\protect\citeauthoryear{Przymusinski}{Przymusinski}{1990}]{Pmodels}
{\sc Przymusinski, T.~C.} 1990.
\newblock The well-founded semantics coincides with the three-valued stable
  semantics.
\newblock {\em Fundam. Inform.\/}~{\em 13,\/}~4, 445--463.

\bibitem[\protect\citeauthoryear{Sato}{Sato}{2017}]{linalg_datalog}
{\sc Sato, T.} 2017.
\newblock A linear algebraic approach to {Datalog} evaluation.
\newblock {\em Theory Pract. Log. Program.\/}~{\em 17,\/}~3, 244--265.

\bibitem[\protect\citeauthoryear{Sato, Sakama, and Inoue}{Sato
  et~al\mbox{.}}{2020}]{linalg_3val}
{\sc Sato, T.}, {\sc Sakama, C.}, {\sc and} {\sc Inoue, K.} 2020.
\newblock From 3-valued semantics to supported model computation for logic
  programs in vector spaces.
\newblock In {\em Proceedings of the 12th International Conference on Agents
  and Artificial Intelligence, {ICAART} 2020}. 758--765.

\bibitem[\protect\citeauthoryear{Seo, Guo, and Lam}{Seo
  et~al\mbox{.}}{2015}]{Socialite}
{\sc Seo, J.}, {\sc Guo, S.}, {\sc and} {\sc Lam, M.~S.} 2015.
\newblock Socialite: An efficient graph query language based on {Datalog}.
\newblock {\em {IEEE} Trans. Knowl. Data Eng.\/}~{\em 27,\/}~7, 1824--1837.

\bibitem[\protect\citeauthoryear{Swift and Warren}{Swift and
  Warren}{2012}]{XSB}
{\sc Swift, T.} {\sc and} {\sc Warren, D.~S.} 2012.
\newblock {XSB}: Extending {Prolog} with tabled logic programming.
\newblock {\em Theory and Practice of Logic Programming\/}~{\em 12,\/}~1-2,
  157–187.

\bibitem[\protect\citeauthoryear{Swift, Warren, et~al\mbox{.}}{Swift
  et~al\mbox{.}}{2017}]{XSB2}
{\sc Swift, T.}, {\sc Warren, D.~S.}, {\sc et~al\mbox{.}} 2017.
\newblock {The XSB System, Version 3.8.x, Volume 1: Programmer’s Manual}.
\newblock Tech. rep.

\bibitem[\protect\citeauthoryear{Tachmazidis and Antoniou}{Tachmazidis and
  Antoniou}{2013}]{strat_bigdata}
{\sc Tachmazidis, I.} {\sc and} {\sc Antoniou, G.} 2013.
\newblock Computing the stratified semantics of logic programs over big data
  through mass parallelization.
\newblock In {\em {RuleML}}. Lecture Notes in Computer Science, vol. 8035.
  Springer, 188--202.

\bibitem[\protect\citeauthoryear{Tachmazidis, Antoniou, and Faber}{Tachmazidis
  et~al\mbox{.}}{2014}]{WFS_bigdata}
{\sc Tachmazidis, I.}, {\sc Antoniou, G.}, {\sc and} {\sc Faber, W.} 2014.
\newblock Efficient computation of the well-founded semantics over big data.
\newblock {\em Theory Pract. Log. Program.\/}~{\em 14,\/}~4-5, 445--459.

\bibitem[\protect\citeauthoryear{Tachmazidis, Antoniou, Flouris, Kotoulas, and
  McCluskey}{Tachmazidis et~al\mbox{.}}{2012}]{ECAI2012}
{\sc Tachmazidis, I.}, {\sc Antoniou, G.}, {\sc Flouris, G.}, {\sc Kotoulas,
  S.}, {\sc and} {\sc McCluskey, L.} 2012.
\newblock Large-scale parallel stratified defeasible reasoning.
\newblock In {\em {ECAI} 2012 - 20th European Conference on Artificial
  Intelligence}. 738--743.

\bibitem[\protect\citeauthoryear{Tachmazidis, Cheng, Kotoulas, Antoniou, and
  Ward}{Tachmazidis et~al\mbox{.}}{2014}]{WFS_X10}
{\sc Tachmazidis, I.}, {\sc Cheng, L.}, {\sc Kotoulas, S.}, {\sc Antoniou, G.},
  {\sc and} {\sc Ward, T.~E.} 2014.
\newblock Massively parallel reasoning under the well-founded semantics using
  {X10}.
\newblock In {\em 26th {IEEE} International Conference on Tools with Artificial
  Intelligence, {ICTAI} 2014}. {IEEE} Computer Society, 162--169.

\bibitem[\protect\citeauthoryear{Tamaki and Sato}{Tamaki and Sato}{1984}]{TS}
{\sc Tamaki, H.} {\sc and} {\sc Sato, T.} 1984.
\newblock Unfold/fold transformation of logic programs.
\newblock In {\em Proc. ICLP}. 127--138.

\bibitem[\protect\citeauthoryear{{Van~Gelder}, Ross, and Schlipf}{{Van~Gelder}
  et~al\mbox{.}}{1991}]{WF91}
{\sc {Van~Gelder}, A.}, {\sc Ross, K.~A.}, {\sc and} {\sc Schlipf, J.~S.} 1991.
\newblock The well-founded semantics for general logic programs.
\newblock {\em J. ACM\/}~{\em 38,\/}~3, 620--650.

\bibitem[\protect\citeauthoryear{Wan, Grosof, Kifer, Fodor, and Liang}{Wan
  et~al\mbox{.}}{2009}]{LPDA}
{\sc Wan, H.}, {\sc Grosof, B.~N.}, {\sc Kifer, M.}, {\sc Fodor, P.}, {\sc and}
  {\sc Liang, S.} 2009.
\newblock Logic programming with defaults and argumentation theories.
\newblock In {\em ICLP}, {P.~M. Hill} {and} {D.~S. Warren}, Eds. Lecture Notes
  in Computer Science, vol. 5649. Springer, 432--448.

\bibitem[\protect\citeauthoryear{Wang, Balazinska, and Halperin}{Wang
  et~al\mbox{.}}{2015}]{MyriaL}
{\sc Wang, J.}, {\sc Balazinska, M.}, {\sc and} {\sc Halperin, D.} 2015.
\newblock Asynchronous and fault-tolerant recursive {Datalog} evaluation in
  shared-nothing engines.
\newblock {\em Proc. {VLDB} Endow.\/}~{\em 8,\/}~12, 1542--1553.

\bibitem[\protect\citeauthoryear{Wang, Hussain, Zuo, Xu, and Sani}{Wang
  et~al\mbox{.}}{2017}]{Graspan}
{\sc Wang, K.}, {\sc Hussain, A.}, {\sc Zuo, Z.}, {\sc Xu, G.~H.}, {\sc and}
  {\sc Sani, A.~A.} 2017.
\newblock Graspan: {A} single-machine disk-based graph system for
  interprocedural static analyses of large-scale systems code.
\newblock In {\em Proceedings of the Twenty-Second International Conference on
  Architectural Support for Programming Languages and Operating Systems,
  {ASPLOS}}. 389--404.

\bibitem[\protect\citeauthoryear{Wenzel and Brass}{Wenzel and
  Brass}{2019}]{embedded}
{\sc Wenzel, M.} {\sc and} {\sc Brass, S.} 2019.
\newblock Declarative programming for microcontrollers - {Datalog on Arduino}.
\newblock In {\em Declarative Programming and Knowledge Management - Conference
  on Declarative Programming, {DECLARE} 2019, Unifying INAP, WLP, and WFLP},
  {P.~Hofstedt}, {S.~Abreu}, {U.~John}, {H.~Kuchen}, {and} {D.~Seipel}, Eds.
  Lecture Notes in Computer Science, vol. 12057. Springer, 119--138.

\bibitem[\protect\citeauthoryear{Whaley, Avots, Carbin, and Lam}{Whaley
  et~al\mbox{.}}{2005}]{bddbddb}
{\sc Whaley, J.}, {\sc Avots, D.}, {\sc Carbin, M.}, {\sc and} {\sc Lam, M.~S.}
  2005.
\newblock Using {Datalog} with binary decision diagrams for program analysis.
\newblock In {\em Programming Languages and Systems, Third Asian Symposium,
  {APLAS}}. 97--118.

\bibitem[\protect\citeauthoryear{Wu, Diamos, Sheard, Aref, Baxter, Garland, and
  Yalamanchili}{Wu et~al\mbox{.}}{2014}]{RedFox}
{\sc Wu, H.}, {\sc Diamos, G.~F.}, {\sc Sheard, T.}, {\sc Aref, M.}, {\sc
  Baxter, S.}, {\sc Garland, M.}, {\sc and} {\sc Yalamanchili, S.} 2014.
\newblock Red fox: An execution environment for relational query processing on
  gpus.
\newblock In {\em 12th Annual {IEEE/ACM} International Symposium on Code
  Generation and Optimization, {CGO}}, {D.~R. Kaeli} {and} {T.~Moseley}, Eds.
  {ACM}, 44.

\bibitem[\protect\citeauthoryear{You and Yuan}{You and Yuan}{1994}]{regular}
{\sc You, J.} {\sc and} {\sc Yuan, L.} 1994.
\newblock A three-valued semantics for deductive databases and logic programs.
\newblock {\em J. Comput. Syst. Sci.\/}~{\em 49,\/}~2, 334--361.

\end{thebibliography}

\appendix

\section{The Metaprogram}  \label{app:metaprogram}

The presentation of $\M_{\pl}$ in the body of the paper sacrifices correct syntax for readability.
In this appendix we present $\M_{\pl}$ in correct logic programming syntax.
This involves a number of auxiliary predicates.

The main clauses of $\M_{\pl}$ are as follows:

\begin{Clause}%\label{strictly1}
  $\mt{definitely}(X)$:-\\
  \> $\mt{fact}(X)$.
\end{Clause}

\begin{Clause}%\label{strictly2}
  $\mt{definitely}(X)$:-\\
  \> $\mt{strict}(R,X,Y)$,\\
  \> $\mt{loop\_definitely}(Y)$.
\end{Clause}

\begin{Clause}%\label{lambda1}
  $\mt{lambda}(X)$:-\\
  \> $\mt{definitely}(X)$.
\end{Clause}

\begin{Clause}%\label{lambda2}
  $\mt{lambda}(X)$:-\\
  \> $\mt{neg}(X, X')$, \\
  \> $\mt{not\ definitely}(X')$,\\
  \> $\mt{strict\_or\_defeasible}(R,X,Y)$,\\
  \> $\mt{loop\_lambda}(Y)$.
\end{Clause}

\begin{Clause}%\label{defeasibly1}
  $\mt{defeasibly}(X)$:-\\
  \> $\mt{definitely}(X)$.
\end{Clause}

\begin{Clause}%\label{defeasibly2}
  $\mt{defeasibly}(X)$:-\\
  \> $\mt{neg}(X, X')$, \\
  \> $\mt{not\ definitely}(X')$,\\
  \> $\mt{strict\_or\_defeasible}(R,X,Y)$,\\
  \> $\mt{loop\_defeasibly}(Y)$,\\
  \> $\mt{not\ overruled}(R,X)$.
\end{Clause}

\begin{Clause}%\label{overruled}
  $\mt{overruled}(R,X)$:-\\
  \> $\mt{neg}(X, X')$, \\
  \> $\mt{rule}(S,X',U)$,\\
  \> $\mt{loop\_lambda}(U)$,\\
  \> $\mt{not\ defeated}(S,X')$.
\end{Clause}

\begin{Clause}%\label{defeated}
  $\mt{defeated}(S,X')$:-\\
  \> $\mt{neg}(X, X')$, \\
  \> $\mt{sup}(T,S)$, \\
  \> $\mt{strict\_or\_defeasible}(T,X,V)$,\\
  \> $\mt{loop\_defeasibly}(V)$.
\end{Clause}

We still need to define the predicates not defined above.
There are additional clauses to represent sets of rules.
\begin{Clause}
  $\mt{rule}(R,H,B)$:-\\
  \> $\mt{strict\_or\_defeasible}(R,H,B)$. \\
\end{Clause}
\begin{Clause}
  $\mt{rule}(R,H,B)$:-\\
  \> $\mt{defeater}(R,H,B)$. \\
\end{Clause}
\begin{Clause}
  $\mt{strict\_or\_defeasible}(R,H,B)$:-\\
  \> $\mt{strict}(R,H,B)$. \\
\end{Clause}
\begin{Clause}
  $\mt{strict\_or\_defeasible}(R,H,B)$:-\\
  \> $\mt{defeasible}(R,H,B)$. \\
\end{Clause}

To express the complement of a literal $\non q$ we define,
for each predicate $p$ in $D$,
\begin{Clause}\label{neg1}
\> $\mt{neg}(p(\ldots), not\_p(\ldots))$. 
\end{Clause}
\begin{Clause}\label{neg2}
\> $\mt{neg}(not\_p(\ldots), p(\ldots))$.
\end{Clause}

The auxiliary predicate $\mt{loop\_defeasibly}$ ($\mt{loop\_lambda}$, $\mt{loop\_definitely}$)
maps the representation of a rule body to a corresponding sequence of calls to 
$\mt{defeasibly}$ (respectively, $\mt{lambda}$, $\mt{definitely}$).

\begin{Clause}\label{loopdefeasibly1}
$\mt{loop\_defeasibly}([]).$ 
\end{Clause}
\begin{Clause}\label{loopdefeasibly2}
$\mt{loop\_defeasibly}([H|T])$ :- $\mt{defeasibly}(H)$, $\mt{loop\_defeasibly}(T)$.
\end{Clause}

\begin{Clause}\label{looplambda1}
$\mt{loop\_lambda}([]).$ 
\end{Clause}
\begin{Clause}\label{looplambda2}
$\mt{loop\_lambda}([H|T])$ :- $\mt{lambda}(H)$, $\mt{loop\_lambda}(T)$.
\end{Clause}
\begin{Clause}\label{loopdefinitely1}
$\mt{loop\_definitely}([]).$ 
\end{Clause}
\begin{Clause}\label{loopdefinitely2}
$\mt{loop\_definitely}([H|T])$ :- $\mt{definitely}(H)$, $\mt{loop\_definitely}(T)$.
\end{Clause}

\end{document}